\documentstyle{article}
\begin{document}
\title{
 Renormalisability of the SU(2)$\times$U(1) Electroweak Theory \\
 with Massive W Z Fields and Massive Matter Fields
\author{Ze-sen Yang, Xianhui Li, Weizhen Deng and Xiaolin Chen \\
Department of Physics,  Peking University, Beijing 100871, CHINA }
%\address{Department of Physics,  Peking University, Beijing 100871, China}
%%\date{March 4, 2000}
}
\date{\today}
\maketitle             
\begin{abstract}
      We extend the previous work and study  the renormalisability of the
 SU$_L$(2) $\times$ U$_Y$(1) electroweak theory with massive W Z fields
 and massive matter fields. We expound that with the constraint conditions
 caused by the W Z mass term and the additional condition chosen by us we
 can still performed the quantization in the same way as before.
 We also show that when the $\delta-$ functions appearing in the path
 integral of the Green functions and representing the constraint conditions
 are rewritten as Fourier integrals with Lagrange multipliers $\lambda_a$
 and $\lambda_y$, the total effective action consisting of the Lagrange
 multipliers, ghost fields and the original fields is BRST invariant.
 Furthermore, with the help of the the renormalisability  of the theory 
 without the the mass term of matter fields, we find the general form of
 the divergent part of the generating functional for the regular vertex
 functions and prove the renormalisability of the theory with the mass terms
 of the W Z fields and the matter fields.
\end{abstract}
\par
\newpage
\begin{center}
{\bf I}.\ \ Introduction
\end{center} \par \ \par
      Owing to the lack of experimental evidence for the Higgs Bosons and
to the unsatisfying treatment on quantization the non-Abelian theory with
massive gauge fields has been reinvestigated [1-9]. Particularly, it has
been clarified [1-3] that the SU(n) theory with massive gauge fields
and the SU(2)$\times$U(1) theory of S.L.Glashow [10] with massive W Z
fields are renormalisable. In the present paper we will extend these work
and study the renormalisability of the latter electroweak theory with 
the massive W Z fields and massive matter fields. For the sake of
convenience we assume that the matter fields consist only of the electron
and electron-neutrino fields. \par
    With the W Z mass term and the matter field mass term directly added to
the Lagrangian by hand, the classical equations of motion will yield
complicated constraint conditions containing products of the field functions.
Moreover, when the constraints are expressed so that the gauge field parts
contain no mass parameters the matter field parts will have a negative
dimension coefficient $m/M^2$, where $M$ and $m$ are the mass parameters of
the W fields and the electron fields respectively. \par
    As in the case of Ref. [3], since the mass term is invariant under an
infinitesimal gauge transformation  with
$\delta \theta_1$ and $\delta \theta_2$ equal to zero and $\delta \theta_3$
equal to $\delta \theta_y$, where $\theta_a$ and $\theta_1$ are the
parameters of the gauge group, an additional constraint condition should be
properly chosen. We will expound that with the constraint conditions caused
by the mass term and the additional condition chosen by us we can performed
the quantization and construct the ghost action in a way similar to that used
in Refs. [1,3]. We will also show that when the $\delta-$ functions appearing in
the path integral of the Green functions and representing the constraint
conditions are rewritten as Fourier integrals with Lagrange multipliers
$\lambda_a$ and $\lambda_y$, the total effective action consisting of the
Lagrange multipliers, ghost fields and the original fields is BRST invariant.
A special thing is that the effective action has a matter--ghost term coming
from the matter field parts of the constraint conditions and containing the
factor $m/M^2$. \par
      We will follow the procedure of Ref. [3] and use the generalized form
of the theory containing $\lambda_a$, $\lambda_y$ and their sources in the
generating functional for the Green functions to study the renormalisability
of the theory containing only the original fields and the ghost fields.
Namely, after deriving the Slavnov--Taylor identities and the additional
identities for the generating functional $\Gamma$ for the regular vertex
functions with the help of the generalized form of the theory, we will let
vanish the functional derivatives of $\Gamma$ with respect to the classical
fields of these Lagrange multipliers. In this way the divergent part of
$\Gamma$ will be shown to satisfy a set of equations which can still be
treated. Furthermore, with the help of the the renormalisability of
the theory without the the mass term of matter fields, we will be able to
find the general form of the divergent part of $\Gamma$ and prove that the
mass term of the matter fields is also harmless to the renormalisability of
the theory. \par
    In spite of the extra complexeity caused by the mass term of the matter
fields we will write this paper in the similar form as that of Ref. [3]. In
section $2$ we will find the constraint conditions coming from the W Z mass
term and choose the additional constraint condition. The method of
quantization will be explained in section $3$. Setion $4$ is devoted to
prove the renormalisability of the theory. Concluding remarks will be given
in the final section.
\par
\vspace{5mm}
\def\theequation{2.\arabic{equation}}
\begin{center}
{\bf II}.\ \ Original and Additional Constraint Conditions
\end{center} \par \ \par
    The matter fields will be often denoted by $\psi(x)$ and
$\overline{\psi}(x)$ and they only contain the electron fields and
electron-neutrino fields in the present work. The former stands for the
purely left-handed neutrino field $\nu_L$, the left- and right-handed parts
of the electron field namely $e_L$, $e_R$, and the latter stands for
$\overline{\nu}_L$, $\overline{e}_L$ and $\overline{e}_R$. Therefore the
mass term of the matter fields is
\begin{eqnarray}
{\cal L}_{\psi m}(x) = - m \overline{e}_L(x) {e}_R(x)
                         - m \overline{e}_R(x) e_L(x)  \,.
\end{eqnarray}
Next let $W_{a \mu}(x)$, $W_{y \mu}(x)$ be the SU$_L$(2) and U$_Y$(1) gauge
fields and $g$, $g_1$ be the coupling constants. Thus the W Z mass term in
the Lagrangian is
\begin{eqnarray}
{\cal L}_{WM} = \frac{1}{2} M^2 W_{a \mu}W_{a}^{\mu} 
          + \frac{1}{2} M^2 \Big(\frac{g_1}{g}\Big)^2 W_{y \mu}W_{y}^{\mu}
          - M^2 \Big(\frac{g_1}{g}\Big) W_{3 \mu}W_{y}^{\mu} \,,
\end{eqnarray}
or
$$
 {\cal L}_{WM} = \frac{1}{2} M^2 W_{1 \mu}(x) W_{1}^{\mu}(x)
                 + \frac{1}{2} M^2 W_{2 \mu}(x) W_{2}^{\mu}(x)
                 + \frac{1}{2} M_z^2 Z_{\mu}(x) Z^{\mu}(x) \,, 
$$ 
where $M_z^2$ stands for $g^{-2}( g^2+ g_1^2) M^2$, and $Z_{\mu}(x)$,
$A_{\mu}(x)$ are the field functions of Z boson and photon, namely 
\begin{eqnarray} 
&& Z_{\mu} =\frac{1}{\sqrt{( g^2+ g_1^2)}} 
  (gW_{3 \mu} - g_1 W_{y \mu})\,, \\
&& A_{\mu} =\frac{1}{\sqrt{( g^2+ g_1^2)}} 
  \varepsilon (g_1 W_{3 \mu} + g W_{y \mu}) \,,
\end{eqnarray}
where $\varepsilon $ is $1$ or $-1$.
\par
    The original Lagrangian of the SU$_L$(2) $\times$ U$_Y$(1) electroweak
theory with the mass term ${\cal L}_{WM}$ is
\begin{eqnarray} 
{\cal L} = {\cal L}_{\psi m} + {\cal L}_{\psi} + {\cal L}_{\psi W}
            + {\cal L}_{WM} + {\cal L}_{WL} + {\cal L}_{WY} \,,
\end{eqnarray} 
where ${\cal L}_{\psi}$ describe the pure matter fields,
${\cal L}_{\psi W}$ is the coupling term between the matter and gauge
fields. ${\cal L}_{WL}$ and ${\cal L}_{WY}$ are the gauge field parts 
without mass terms, namely
\begin{eqnarray} 
&&{\cal L}_{WL} = -\frac{1}{4} F_{a \mu \nu} F_a^{\mu \nu} \,,\\
&&{\cal L}_{WY} = -\frac{1}{4} B_{\mu \nu} B^{\mu \nu} \,,
\end{eqnarray} 
where              
\begin{eqnarray} 
&& F_{a \mu \nu} = \partial_{\mu}W_{a \nu}-\partial_{\nu}W_{a \mu}
- g C_{abc} W_{b \mu } W_{c \nu} \,,\\
&& B_{\mu \nu} = \partial_{\mu}W_{y \nu}-\partial_{\nu}W_{y \mu}\,.
\end{eqnarray} 
\par
\noindent
$C_{abc}$ stands for the structure constants of SU$_L$(2) with $C_{123}$
equal to $1$.
\par
    Denote by $ \theta_a(x), \theta_y(x) $ the parameters of the gauge group.
Thus, under an infinitesimal gauge transformation, the fields $W_a^{\mu}$,
$W_y^{\mu}$, $\psi$ and $\overline{\psi}$ transform as
\begin{eqnarray*}  
&& \delta W_a^{\mu}(x) =
    - \frac{1}{g} \partial^{\mu} \delta \theta_a(x)
     - C_{abc} W_c^{\mu}(x) \delta \theta_b(x)  \,,\\
&& \delta W_y^{\mu}(x) =
    - \frac{1}{g_1} \partial^{\mu} \delta \theta_y(x) \,,\\
&& \delta \nu_L(x) = \frac{i}{2}\delta \theta_1(x) e_L(x)
+ \frac{1}{2}\delta \theta_2(x)e_L(x)+ \frac{i}{2}\delta \theta_3(x)\nu_L(x)
   - \frac{i}{2}\delta \theta_y(x) \nu_L(x)\,,\\
&& \delta e_L(x) = \frac{i}{2}\delta \theta_1(x) \nu_L(x)
 - \frac{1}{2}\delta \theta_2(x)\nu_L(x) -\frac{i}{2}\delta \theta_3(x)e_L(x)
 - \frac{i}{2}\delta \theta_y(x)e_L(x) \,,\\
&& \delta e_R(x) = -i \delta \theta_y(x) e_R(x) \,,\\
&&\delta \overline{\nu}_L(x) = -\frac{i}{2}\delta\theta_1(x)\overline{e}_L(x)
    + \frac{1}{2}\delta \theta_2(x) \overline{e}_L(x)
    - \frac{i}{2}\delta \theta_3(x) \overline{\nu}_L(x)
    + \frac{i}{2}\delta \theta_y(x) \overline{\nu}_L(x)\,,\\
&&\delta \overline{e}_L(x)= -\frac{i}{2}\delta \theta_1(x)\overline{\nu}_L(x)
    - \frac{1}{2}\delta \theta_2(x)\overline{\nu}_L(x)
    + \frac{i}{2} \delta \theta_3(x) \overline{e}_L(x)
    + \frac{i}{2}\delta \theta_y(x) \overline{e}_L(x) \,,\\
&&\delta \overline{e}_R(x) = i \delta \theta_y(x) \overline{e}_R(x)\,.
\end{eqnarray*} 
$\delta {\cal L}_{\psi m}$ can be written as
\begin{eqnarray} 
\delta {\cal L}_{\psi m}
     = f_a(x) \delta \theta_a(x) + f_y(x) \delta \theta_y(x) \,,
\end{eqnarray} 
where
\begin{eqnarray} 
&& f_1(x) = \frac{i}{2}m \Big\{ \overline{\nu}_L(x) e_R(x)
          - \overline{e}_R(x) \nu_L(x) \Big\} \,,\\
&& f_2(x) = \frac{1}{2}m \Big\{ \overline{\nu}_L(x) e_R(x)
          + \overline{e}_R(x) \nu_L(x) \Big\} \,,\\
&& f_3(x) = \frac{i}{2}m \Big\{ \overline{e}_R(x) e_L(x)
          - \overline{e}_L(x) e_R(x) \Big\} \,,\\
&& f_y(x) = - f_3(x) \,.
\end{eqnarray} 
Therefore the action transforms as
\begin{eqnarray} 
&& \delta \int d^4x {\cal L}(x) =
  \delta \int d^4x \Big\{ {\cal L}_{WM}(x) + {\cal L}_{\psi m}(x) \Big\} 
\nonumber \\
&&\ \ \ \ \ \ \ \ =\,\int d^4x\Big\{
 \Big( \frac{M^2}{g} \partial_{\mu} W_1^{\mu}(x)
       + \frac{M^2}{g}g_1 W_{2 \mu}(x) W_y^{\mu}(x) + f_1(x) 
       \Big) \delta \theta_1 \nonumber\\
&&\ \ \ \ \ \ \ \ \ \ \ \
 + \Big( \frac{M^2}{g} \partial_{\mu} W_2^{\mu}(x)
        - \frac{M^2}{g}g_1 W_{1 \mu}(x) W_y^{\mu}(x) + f_2(x) 
        \Big) \delta \theta_2 \nonumber\\
&&\ \ \ \ \ \ \ \ \ \ \ \
 + \Big( \frac{M^2}{g} \partial_{\mu} W_3^{\mu}(x)
         - \frac{M^2}{g^2} g_1 \partial_{\mu}W_y^{\mu}(x) + f_3(x) 
     \Big) ( \delta \theta_3 - \delta \theta_y )   \Big\} \,.
\end{eqnarray} 
Since the classical equations of motion make the action invariant under
an arbitrary infinitesimal transformation of the field functions, they 
certainly make the mass term invariant under an arbitrary infinitesimal
gauge transformation. This means that when $M$ is not equal to zero, the
classical equations of motion leads to the following constraint conditions 
\begin{eqnarray}
&& \frac{M^2}{g} \partial_{\mu} W_1^{\mu}(x)
      + \frac{M^2}{g}g_1 W_{2 \mu}(x) W_y^{\mu}(x) + f_1(x) = 0 \,,\\
&& \frac{M^2}{g} \partial_{\mu} W_2^{\mu}(x)
      - \frac{M^2}{g}g_1 W_{1 \mu}(x) W_y^{\mu}(x) + f_2(x) = 0 \,,\\
&& \frac{M^2}{g} \partial_{\mu} W_3^{\mu}(x)
       - \frac{M^2}{g^2} g_1 \partial_{\mu}W_y^{\mu}(x) + f_3(x) = 0 \,.
\end{eqnarray}
These are the original constraint conditions. Since the mass term is
invariant under an infinitesimal gauge
transformation with $\delta \theta_1$ and $\delta \theta_2$ equal to zero
and $\delta \theta_3$ equal to $\delta \theta_y$,  
$\partial_{\mu}W_3^{\mu} $ and $\partial_{\mu}W_y^{\mu} $ appear in one
constraint. We now choose an additional condition and replace (2.18) with
\begin{eqnarray}
&& \frac{M^2}{g} \partial_{\mu} W_3^{\mu}(x)
       + \frac{M^2}{g} g_1 W_{3 \mu}(x)W_y^{\mu}(x) + f_3(x) = 0 \,,\\
&& \partial_{\mu}W_y^{\mu}(x) + g W_{3 \mu}(x) W_y^{\mu}(x) = 0 \,.
\end{eqnarray}
\par
\vspace{5mm}
\def\theequation{3.\arabic{equation}}
\begin{center}
{\bf III}.\ \ Quantization and BRST Invariance
\end{center} \par \ \par
\setcounter{equation}{0}
    Write (2.16), (2.17) and (2.19),(2.20) as
\begin{eqnarray}
 \Phi_a(x) = 0 \,,\ \ \ \ \ \  \Phi_y(x) = 0 \,,
\end{eqnarray}
with
\begin{eqnarray}
&& \Phi_1(x) = \partial_{\mu} W_1^{\mu}(x)
             + g_1 W_{2 \mu}(x) W_y^{\mu}(x) + \frac{g}{M^2} f_1(x)  \,,\\
&& \Phi_2(x) = \partial_{\mu} W_2^{\mu}(x)
             - g_1 W_{1 \mu}(x) W_y^{\mu}(x) + \frac{g}{M^2} f_2(x)  \,,\\
&& \Phi_3(x) = \partial_{\mu} W_3^{\mu}(x)
             + g_1 W_{3 \mu}(x)W_y^{\mu}(x) + \frac{g}{M^2} f_3(x)  \,,\\
&& \Phi_y(x) = \partial_{\mu}W_y^{\mu}(x)
             + g W_{3 \mu}(x) W_y^{\mu}(x)  \,.
\end{eqnarray}
\par
    Taking the constraint conditions (3.1) into account one should write
the path integral of the Green functions inolving only the original fields
as
\begin{eqnarray} 
 \frac{1}{N_0} \int {\cal D}[{\cal W},\overline{\psi},\psi] 
 \Delta [{\cal W},\overline{\psi},\psi]
\prod_{a',x'} \delta\left(\Phi_{a'}(x')\right)\delta\left(\Phi_y(x') \right)
W_{a \mu} (x)W_{b \nu}(y) \cdots {\rm exp} \{{\rm i} I \} \,,
\end{eqnarray} 
where
\begin{eqnarray*} 
&& I = \int d^4x {\cal L}(x) \,,\\
&& N_0 =  \int {\cal D}[{\cal W},\overline{\psi},\psi]
 \Delta [{\cal W},\overline{\psi},\psi]
\prod_{a',x'} \delta \left(\Phi_{a'}(x')\right)\delta\left(\Phi_y(x') \right) 
 {\rm exp} \{{\rm i} I \} \,.
\end{eqnarray*}
Since only the field functions which satisfy the constraint
conditions can play roles in the integral (3.6), the value of the Lagrangian
can be changed for the field functions which do not satisfy these conditions.
In view of the fact that the conditions (3.1) make the action invariant with
respect to the infinitesimal gauge trasformation, we now imagine to replace
the mass term ${\cal L}_{WM}$ in (3.6) with a gauge invariant mass term
which is equal to ${\cal L}_{WM}$ when the conditions (3.1) are satisfied.
Thus, analogous to the case in the Fadeev--Popov method [1,3,11-16],
$\Delta[{\cal W},\overline{\psi},\psi]$ should be gauge invariant and make
the following equation valid for an arbitrary gauge invariant quantity
${\cal O}({\cal W,\,\overline{\psi},\,\psi })$
\begin{eqnarray*}
&&  \int {\cal D}[{\cal W},\,\overline{\psi},\,\psi] 
 \Delta[{\cal W},\overline{\psi},\psi]
 \prod_{a',x'}
 \delta \left(\Phi_{a'}(x') \right)\delta\left(\Phi_y(x') \right)
 {\cal O}({\cal W},\overline{\psi},\psi)
{\rm exp} \{{\rm i} \widetilde{I} \} \\
&&\ \ \ \ \ \ \ \ \ \ \ \
 \propto  \int {\cal D}[{\cal W},\,\overline{\psi},\,\psi]
{\cal O}({\cal W},\overline{\psi},\psi)
{\rm exp} \{{\rm i} \widetilde{I} \}   \,,
\end{eqnarray*}
where $\widetilde{I}$ is a gauge invariant action constructed by replacing
${\cal L}_{WM}$ with the imagined mass term. This means that the weight
factor $\Delta[{\cal W},\overline{\psi},\psi]$ can be determined according
to the Fadeev--Popov equation of the following form
\begin{eqnarray}
\Delta[{\cal W},\overline{\psi},\psi]
\int \prod_z d \Omega (z) \prod_{\sigma,x} \delta
\Big( \Phi_{\sigma}^{\Omega}(x) \Big) = 1 \,.
\end{eqnarray}
where $\sigma$ stands for $1,2,3,y$, $\Phi_{\sigma}^{\Omega}(x)$ is the
result of acting on $\Phi_{\sigma}(x)$ with a gauge transformation having
the parameters of the element $\Omega(x)$ of the gauge group, $d\Omega (z)$
is the volume element of the group integral. It follows that with the F--P
ghost fields $C_a(x)$, $C_y(x)$, $\overline{C}_a(x)$, $\overline{C}_y(x)$
as new variables, one can express the ghost Lagrangian as
\begin{eqnarray}
 {\cal L}^{(C)}(x) = \overline{C}_a(x) \Delta \Phi_a(x)
                      + \overline{C}_y(x) \Delta \Phi_y(x)\,,
\end{eqnarray}
where $\Delta \Phi_a(x)$, $\Delta \Phi_y(x)$ are defined by the BRST
transformtion of $\Phi_a(x)$ and $\Phi_y(x)$ so that
\begin{eqnarray}
\delta_B \Phi_a(x) = \delta \zeta \Delta \Phi_a(x) \,,\ \ \ \ 
  \delta_B \Phi_y(x) = \delta \zeta \Delta \Phi_y(x) \,,                     
\end{eqnarray}
where $\delta \zeta $ is an infinitesimal fermionic parameter independent
of $x$. The BRST transformation of the gauge fields or matter fields is
nothing but the infinitesimal gauge transformation with $\delta \theta_a$
and $\delta \theta_y$ equal to $-g \delta \zeta C_a $
and $-g_1 \delta \zeta C_y $ respectively. Namely
\begin{eqnarray}
&& \delta_B W_a^{\mu}(x)  = \delta \zeta \Delta W_a^{\mu}(x) 
                       = \delta \zeta D^{\mu}_{ab}C_b(x) \,,\\
&& \delta_B W_y^{\mu}(x) = \delta \zeta \Delta W_y^{\mu}(x) 
                      = \delta \zeta \partial^{\mu} C_y(x) \,,\\
&& \delta_B \psi(x) = \delta \zeta \Delta \psi(x) \,, \ \ \ \ \ 
  \delta_B \overline{\psi}(x) = \delta \zeta \Delta \overline{\psi}(x) \,,
\end{eqnarray}
where
\begin{eqnarray*}
&& D^{\mu}_{ab}(x) =  \delta_{ab} \partial^{\mu}
                     + g f_{abc} A_c^{\mu}(x)  \,, \\
&& \Delta \nu_L(x) = - \frac{i}{2} gC_1(x) e_L(x)
  - \frac{1}{2} gC_2(x)e_L(x) - \frac{i}{2} gC_3(x)\nu_L(x)
   + \frac{i}{2} g_1 C_y(x) \nu_L(x)\,,\\
&& \Delta e_L(x) = - \frac{i}{2} gC_1(x) \nu_L(x)
   + \frac{1}{2} gC_2(x)\nu_L(x) + \frac{i}{2} gC_3(x)e_L(x)
   + \frac{i}{2} g_1 C_y(x)e_L(x) \,,\\
&& \Delta e_R(x) = i g_1 C_y(x) e_R(x) \,,\\
&&\Delta \overline{\nu}_L(x) = \frac{i}{2} gC_1(x)\overline{e}_L(x)
    - \frac{1}{2} gC_2(x) \overline{e}_L(x)
    + \frac{i}{2} gC_3(x) \overline{\nu}_L(x)
    - \frac{i}{2} g_1 C_y(x) \overline{\nu}_L(x)\,,\\
&&\Delta \overline{e}_L(x) = \frac{i}{2} gC_1(x)\overline{\nu}_L(x)
    + \frac{1}{2} gC_2(x)\overline{\nu}_L(x)
    - \frac{i}{2} gC_3(x) \overline{e}_L(x)
    - \frac{i}{2} g_1 C_y(x) \overline{e}_L(x) \,,\\
&&\Delta \overline{e}_R(x) = -i g_1 C_y(x) \overline{e}_R(x)\,.
\end{eqnarray*}
 $ C_a(x)$ and $C_y(x) $ are also transformed as usual
\begin{eqnarray*}
&& \delta_B C_a(x) = \delta \zeta \Delta C_a(x) 
                = \delta \zeta \frac{g}{2} C_{abc}C_b(x)C_c(x) \,,\\
&& \delta_B C_y(x) = 0 \,.
\end{eqnarray*}
Now we can write $\Delta \Phi_a(x)$, $ \Delta \Phi_y(x)$ as
\begin{eqnarray}
&&  \Delta \Phi_1 = \partial_{\mu}\Delta W_1^{\mu}(x)
                  + g_1 \Delta W_2^{\mu}(x)  W_{y \mu}(x)
                  + g_1 W_{2 \mu}(x)\Delta W_y^{\mu}(x)
                  + \frac{g}{M^2} \Delta f_1(x) \,, \ \ \  \\
&&  \Delta \Phi_2 =  \partial_{\mu}\Delta W_2^{\mu}(x)
                  - g_1 \Delta W_1^{\mu}(x)  W_{y \mu}(x)
                  - g_1 W_{1 \mu}(x)\Delta W_y^{\mu}(x)
                  + \frac{g}{M^2} \Delta f_2(x) \,, \ \ \  \\
&&  \Delta \Phi_3 = \partial_{\mu}\Delta W_3^{\mu}(x)
                  + g_1 \Delta W_3^{\mu}(x)  W_{y \mu}(x)
                  + g_1 W_{3 \mu}(x)\Delta W_y^{\mu}(x)
                  + \frac{g}{M^2} \Delta f_3(x) \,, \ \ \  \\
&&  \Delta \Phi_y = \partial_{\mu}\Delta W_y^{\mu}(x)
                  + g \Delta W_3^{\mu}(x) W_{y \mu}(x)
                  + g W_{3 \mu}(x)\Delta W_y^{\mu}(x) \,, \ \ \ 
\end{eqnarray}
where
\begin{eqnarray*}
&&\Delta f_1(x) = \frac{i}{2}m \Big\{
                 \Big(\Delta \overline{\nu}_L(x)\Big) e_R(x)
                 - \overline{\nu}_L(x) \Delta e_R(x)
                 - \Big(\Delta \overline{e}_R(x)\Big) \nu_L(x)
                 + \overline{e}_R(x) \Delta \nu_L(x)  \Big\} \,,\\
&&\Delta f_2(x) = \frac{1}{2}m \Big\{
                \Big(\Delta \overline{\nu}_L(x)\Big) e_R(x)
                - \overline{\nu}_L(x) \Delta e_R(x)
                + \Big( \Delta \overline{e}_R(x)\Big) \nu_L(x)
                - \overline{e}_R(x) \Delta \nu_L(x) \Big\} \,,\\
&& \Delta f_3(x) = \frac{i}{2}m \Big\{
                 \Big( \Delta \overline{e}_R(x)\Big) e_L(x)
                 - \overline{e}_R(x) \Delta e_L(x)
                 - \Big( \Delta \overline{e}_L(x)\Big) e_R(x)
                 + \overline{e}_L(x) \Delta e_R(x) \Big\} \,.
\end{eqnarray*}
Since $\Delta W_a^{\mu}$, $\Delta W_y^{\mu}$, $\Delta\psi(x)$,
 $\Delta\overline{\psi}(x)$ and $\Delta C_a(x)$ are BRST invariant, it is
easy to see that $\Delta \Phi_a(x)$ and $\Delta \Phi_y(x)$ are also BRST
invariant.
\par
    One can further generalized the theory by regarding as new variables
the Lagrange multipliers $\lambda_a(x)$ and $\lambda_y(x)$ associated with
the constraint conditions. Thus the total effective Lagrangian and action
consist of these Lagrange multipliers, ghosts and the original variables, 
namely
\begin{eqnarray}
&&{\cal L}_{{\rm eff}}(x) = {\cal L}(x) + {\cal L}^{(C)}(x)
                       + \lambda_a(x) \Phi_a(x) 
                       + \lambda_y(x) \Phi_y(x)  \,,\\
&& I_{{\rm eff}} = \int d^4x {\cal L}_{{\rm eff}}(x)  \,. 
\end{eqnarray}
Correspondingly, the path integral of the generating functional for the
Green functions is
\begin{eqnarray}
 {\cal Z}[\overline{\eta}, \eta,\overline{\chi},\chi,J,j]
= \frac{1}{N_{\lambda}}
\int {\cal D}[\overline{\psi},\psi,{\cal W}, \overline{C},C,\lambda]
 {\rm exp} \Big\{ {\rm i} \big( I_{{\rm eff}} + I_s \big) \Big\},
\end{eqnarray}
where $N_{\lambda}$ is a constant, $I_s$ is the source term in the action.
They are defined by
\begin{eqnarray}  
&& N_{\lambda}
    = \int {\cal D}[\overline{\psi}, \psi,{\cal W},\overline{C},C,\lambda]
     {\rm exp} \Big\{ {\rm i} I_{{\rm eff}}  \Big\} \,, \nonumber\\
&&I_s = \int d^4x \Big\{\overline{\eta}(x)\psi(x) + \overline{\psi}(x)\eta(x)  
          + \overline{\chi}_a(x)C_a(x) + \overline{C}_a(x) \chi_a(x)  
          + \overline{\chi}_y(x)C_y(x) \nonumber\\
&&\ \ \ \ \ \ \ \ 
          + \overline{C}_y(x) \chi_y(x)
          + J_a^{\mu}(x) W_{a \mu}(x) + J_y^{\mu}(x) W_{y \mu}(x)
          + j_a(x) \lambda_a(x) + J_y(x) \lambda_y(x)  \Big\} \,,
\end{eqnarray}
where $\overline{\eta}(x),\eta(x)\cdots$ stand for the sources.
In particular, $j_a(x)$, $j_y(x)$ are the sources of $\lambda_a(x)$,
$\lambda_y(x)$, respectively. 
\par
    We now check the BRST invariance of the effective action $I_{eff}$
defined by (3.17) and (3.18). With ${\overline C}_a(x)$, ${\overline C}_y(x)$
transforming as 
$$
 \delta_B {\overline C}_a(x) = - \delta \zeta \lambda_a(x)\,, \ \ \ \ 
 \delta_B {\overline C}_y(x) = - \delta \zeta \lambda_y(x)\,.
$$
and noticing the invariance of $\Delta\Phi_a, \Delta\Phi_y$, one has
$$
 \delta_B \int d^4x {\cal L}^{(C)}(x)
  = \int d^4x \Big\{ -\lambda_a(x) \delta_B \Phi_a(x)
                    - \lambda_y(x) \delta_B \Phi_y(x) \Big\} \,.
$$
Therefore
$$
 \delta_B I_{{\rm eff}} 
  = \delta_B \int d^4x \Big\{ {\cal L}_{WM} + {\cal L}_{\psi m} \Big\}
         + \int d^4x \Big\{ \big( \delta_B\lambda_a(x) \big) \Phi_a(x)
                   + \big( \delta_B \lambda_y(x) \big) \Phi_y(x) \Big\} \,.
$$
From this and the expression of $\delta_B I_{WM}$, it can be shown that
the effective action is invariant, when the transformation of $\lambda_a(x)$
and $\lambda_y(x)$ are defined as
\begin{eqnarray*}
&& \delta_B \lambda_1(x) = \delta \zeta M^2 C_1(x) \,,\\
&& \delta_B \lambda_2(x) = \delta \zeta M^2 C_2(x) \,,\\
&& \delta_B \lambda_3(x)
    = \delta \zeta M^2 C_3(x)
       - \delta \zeta \frac{g_1}{g} M^2 C_y(x)  \,,\\
&& \delta_B \lambda_y(x)
= \delta \zeta \frac{g_1^2}{g^2} M^2 C_y(x)
  - \delta \zeta \frac{g_1}{g} M^2 C_3(x)  \,.
\end{eqnarray*}
\par
\vspace{5mm}
\def\theequation{4.\arabic{equation}}
\begin{center}
{\bf IV}.\ \ Renormalisability 
\end{center} \par \ \par
\setcounter{equation}{0}
    Following the notations of Ref. [3],
    let $W_{a \mu}(x), W_{y \mu}(x) $, $C_a(x), C_y(x), \cdots $ stand
for the renormalized field founctions, $g, g_1$ and $M$ be renormalized
parameters. By introducing the source terms of the composite field
functions  $\Delta W_a^{\mu}$, $\Delta W_y^{\mu}$, $\Delta C_a(x)$,
$\Delta \psi(x)$, $\Delta \overline{\psi}(x)$ and the sources $K^a_{\mu}(x)$,
$K^y_{\mu}(x)$, $L_a(x)$, $n_{\alpha}(x)$, $l_{\alpha}(x)$, $p_{\alpha}(x)$,
 $n'_{\alpha}(x)$, $l'_{\alpha}(x)$ and $p'_{\alpha}(x)$, the effective
Lagrangian without counterterm becomes
\begin{eqnarray}
 {\cal L}^{[0]}_{eff}(x)  
   &=& \lambda_a(x) \Phi_a(x)
       + \lambda_y(x) \Phi_y(x) 
       + {\cal L}_{WL}(x) + {\cal L}_{WY}(x)  \nonumber\\
    &&+ {\cal L}_{WM}(x) + {\cal L}^{(C)}(x) 
      + {\cal L}_{\psi}(x) + {\cal L}_{\psi m}(x) + {\cal L}_{\psi W}(x)
       \nonumber\\
    &&+ K^a_{\mu}(x) \Delta W_a^{\mu}(x)
      + K^y_{\mu}(x) \Delta W_y^{\mu}(x) + L_a(x) \Delta C_a(x) \nonumber\\ 
    &&+ n_{\alpha}(x) \Delta \nu_{L \alpha}(x)
      + l_{\alpha}(x) \Delta e_{L \alpha}(x)
      + p_{\alpha}(x) \Delta e_{R \alpha}(x)\nonumber\\
    &&+ n'_{\alpha}(x) \Delta \overline{\nu}_{L \alpha}(x)
      + l'_{\alpha}(x) \Delta \overline{e}_{L \alpha}(x)
      + p'_{\alpha}(x) \Delta \overline{e}_{R \alpha}(x)   \,.
\end{eqnarray}
The complete effective Lagrangian is the sum of ${\cal L}^{[0]}_{eff}$ and
the counterterm ${\cal L}_{count}$ 
\begin{eqnarray}
{\cal L}_{{\rm eff}} = {\cal L}^{[0]}_{{\rm eff}} + {\cal L}_{count} \,.
\end{eqnarray}
\par
With (4.1), the generating functional for Green functions is defined as
\begin{eqnarray}
{\cal Z}^{[0]}
     [\overline{\eta}, \eta,\overline{\chi},\chi,J,j,K,L,n,l,p,n',l',p']
 = \frac{1}{N} \int {\cal D}
      [\overline{\psi},\psi,{\cal W}, \overline{C},C,\lambda]
   {\rm exp} \Big\{{\rm i} \big( I^{[0]}_{eff} + I_s \big) \Big\}\,, 
\end{eqnarray}
 $I^{[0]}_{eff}$ is the effective action $\int d^4x{\cal L}^{[0]}_{eff}(x)$,
$N$ is a constant to make ${\cal Z}^{[0]}$ equal to $1$ in the absence of
the sources, $I_s$ is the source term
\begin{eqnarray*}
&&I_s = \int d^4x \Big\{\overline{\eta}(x)\psi(x) + \overline{\psi}(x)\eta(x)
          + \overline{\chi}_a(x)C_a(x) + \overline{C}_a(x) \chi_a(x)  
          + \overline{\chi}_y(x)C_y(x) \\
&&\ \ \ \ \ \ \ \ 
          + \overline{C}_y(x) \chi_y(x)
          + J_a^{\mu}(x) W_{a \mu}(x) + J_y^{\mu}(x) W_{y \mu}(x)
          + j_a(x) \lambda_a(x) + j_y(x) \lambda_y(x)  \Big\} \,,
\end{eqnarray*}
where $\overline{\eta}\, \psi$ and $\overline{\psi}\, \eta$ stand for
\begin{eqnarray*}  
&&\overline{\eta}\, \psi = 
        \overline{\eta}^{(\nu)}_{\alpha}\, \nu_{L \alpha}
      + \overline{\eta}^{(l)}_{\alpha}\,  e_{L \alpha}
      + \overline{\eta}^{(r)}_{\alpha}\,  e_{R \alpha} \,,\\
&&\overline{\psi}\, \eta =
        \overline{\nu}_{L \alpha}\, \eta^{(\nu)}_{\alpha}
      + \overline{e}_{L \alpha}\, \eta^{(l)}_{\alpha}
      + \overline{e}_{R \alpha}\, \eta^{(r)}_{\alpha}  \,.
\end{eqnarray*}
Denoting by ${\cal W}^{[0]}$ and $\Gamma^{[0]}$ the generating functionals
for connected Green functions and regular vertex functions respectively,
one has
\begin{eqnarray}
&&{\cal Z}^{[0]} 
= {\rm exp} \Big\{ {\rm i}{\cal W}^{[0]}
 [\overline{\eta},\eta,\overline{\chi},\chi,J,j,K,L,n,l,p,n',l',p'] \Big\}\,,
\\
&& \Gamma^{[0]}
 [\widetilde{\psi},\widetilde{\overline{\psi}},\widetilde{W},
  \widetilde{\overline{C}}, \widetilde{C},
  \widetilde{\lambda},K,L,n,l,p,n',l',p'] \nonumber\\
&&\ \ \ \ \ \ \ \ = {\cal W}^{[0]} 
        - \int d^4x \Big[ J_a^{\mu} \widetilde{W}_{a \mu}
       + J_y^{\mu} \widetilde{W}_{y \mu}
       + j_a \widetilde{\lambda}_a
       + j_y \widetilde{\lambda}_y 
       + \overline{\chi}_a \widetilde{C}_a
       + \widetilde{\overline{C}}_a \chi_a
       + \overline{\chi}_y \widetilde{C}_y  \nonumber\\
&& \ \ \ \ \ \ \ \ \ \ \ \ \ \ \ \ \ \ \
       + \widetilde{\overline{C}}_y \chi_y
       + \overline{\eta}^{(\nu)} \widetilde{\nu}_L
       + \overline{\eta}^{(l)} \widetilde{e}_L 
       + \overline{\eta}^{(r)} \widetilde{e}_R 
       + \widetilde{\overline{\nu}}_L \eta^{(\nu)}
       + \widetilde{\overline{e}}_L \eta^{(l)}
       + \widetilde{\overline{e}}_R \eta^{(r)}   \Big]\,,
\end{eqnarray}
where $\widetilde{W}_{a \mu}$, $\widetilde{\nu}_L$, $\cdots$ are the
so-called classical fields defined by 
\begin{eqnarray*}
&& \widetilde{W}_{a \mu}(x) = \frac{\delta {\cal W}^{[0]}}
                                  {\delta J_a^{\mu}(x) }\,,\ \ \ \ \ \ 
 \widetilde{\lambda}_a(x) = \frac{\delta {\cal W}^{[0]}}
                                  {\delta j_a(x) }\,,\ \ \ \ \ \ 
 \widetilde{C}_a(x) = \frac{\delta {\cal W}^{[0]}}
                                  {\delta \overline{\chi}_a(x)}\,,\\
&& \widetilde{\overline{C}}_a(x) = - \frac{\delta {\cal W}^{[0]}}
                                         {\delta \chi_a(x) }\,,\ \ \ \ \ \
 \widetilde{W}_{y \mu}(x) = \frac{\delta {\cal W}^{[0]}}
                                  {\delta J_y^{\mu}(x) }\,,\ \ \ \ \ \ 
 \widetilde{\lambda}_y = \frac{\delta {\cal W}^{[0]}}
                                  {\delta j_y(x) }\,, \\
&& \widetilde{C}_y(x) = \frac{\delta {\cal W}^{[0]}}
                                  {\delta \overline{\chi}_y(x)}\,,\ \ \ \ \ \
 \widetilde{\overline{C}}_y(x) = - \frac{\delta {\cal W}^{[0]}}
                                         {\delta \chi_y(x) }\,,\ \ \ \ \ \
 \widetilde{\nu}_{L \alpha}(x) = \frac{\delta {\cal W}^{[0]}}
              {\delta \overline{\eta}^{(\nu)}_{\alpha}(x) }\,, \\
&& \widetilde{e}_{L \alpha}(x) = \frac{\delta {\cal W}^{[0]}}
               {\delta \overline{\eta}^{(l)}_{\alpha}(x)}\,,\ \ \ \ \ \
 \widetilde{e}_{R \alpha}(x) = \frac{\delta {\cal W}^{[0]}}
               {\delta \overline{\eta}^{(r)}_{\alpha}(x) }\,,\ \ \ \ \ \
 \widetilde{\overline{\nu}}_{L \alpha}(x) = - \frac{\delta {\cal W}^{[0]}}
                           {\delta \eta^{(\nu)}_{\alpha}(x) }\,, \\
&& \widetilde{\overline{e}}_{L \alpha}(x) = - \frac{\delta {\cal W}^{[0]}}
                           {\delta \eta^{(l)}_{\alpha}(x)}\,,\ \ \ \ \ \
 \widetilde{\overline{e}}_{R \alpha}(x) = - \frac{\delta {\cal W}^{[0]}}
               {\delta \eta^{(r)}_{\alpha}(x) }\,,
\end{eqnarray*}
Therefore
\begin{eqnarray*}
&& J_a^{\mu}(x) = - \frac{\delta \Gamma^{[0]}}
                        {\delta \widetilde{W}_{a \mu}(x) } \,, \ \ \ \ \ \ 
 j_a(x) = - \frac{\delta \Gamma^{[0]}}
                        {\delta \widetilde{\lambda}_a(x) } \,, \ \ \ \ \ \ 
 \overline{\chi}_a(x) = \frac{\delta \Gamma^{[0]}}
                              {\delta \widetilde{C}_a(x)} \,,  \\
&& \chi_a(x) = - \frac{\delta \Gamma^{[0]}}
                     {\delta \widetilde{\overline{C}}_a(x) }\,, \ \ \ \ \ \ 
 J_y^{\mu}(x) = - \frac{\delta \Gamma^{[0]}}
                        {\delta \widetilde{W}_{y \mu}(x) } \,, \ \ \ \ \ \ 
 j_y(x) = - \frac{\delta \Gamma^{[0]}}
                        {\delta \widetilde{\lambda}_y(x) } \,,  \\
&& \overline{\chi}_y(x) = \frac{\delta \Gamma^{[0]}}
                              {\delta \widetilde{C}_y(x)} \,, \ \ \ \ \ \ 
 \chi_y(x) = - \frac{\delta \Gamma^{[0]}}
                     {\delta \widetilde{\overline{C}}_y(x) }\,, \ \ \ \ \ \ 
 {\eta}^{(\nu)}_{\alpha}(x) = - \frac{\delta \Gamma^{[0]}}
              {\delta \widetilde{\overline{\nu}}_{L \alpha}(x) }\,, \\
&&{\eta}^{(l)}_{\alpha}(x)  = - \frac{\delta \Gamma^{[0]}}
               {\delta \widetilde{\overline{e}}_{L \alpha}(x)}\,,\ \ \ \ \ \
{\eta}^{(r)}_{\alpha}(x)  = - \frac{\delta \Gamma^{[0]}}
              {\delta \widetilde{\overline{e}}_{R \alpha}(x) }\,, \ \ \ \ \ \
 \overline{\eta}^{(\nu)}_{\alpha}(x) =  \frac{\delta \Gamma^{[0]}}
              {\delta \widetilde{\nu}_{L \alpha}(x) }\,, \\
&&\overline{\eta}^{(l)}_{\alpha}(x)  =  \frac{\delta \Gamma^{[0]}}
               {\delta \widetilde{e}_{L \alpha}(x)}\,,\ \ \ \ \ \
\overline{\eta}^{(r)}_{\alpha}(x)  =  \frac{\delta \Gamma^{[0]}}
               {\delta \widetilde{e}_{R \alpha}(x) }\,.
\end{eqnarray*}
Besides, for $K^a_{\mu}, L_a$ $\cdots$, the spectators in the Legendre
transtrormation, one has
\begin{eqnarray*}
&& \frac{\delta {\cal W}^{[0]}}{\delta K^a_{\mu}(x) }
       = \frac{\delta \Gamma^{[0]}}{\delta K^a_{\mu}(x) } \,, \ \ \ \ \ \
 \frac{\delta {\cal W}^{[0]}}{\delta K^y_{\mu}(x) }
       = \frac{\delta \Gamma^{[0]}}{\delta K^y_{\mu}(x) } \,, \ \ \ \ \ \
 \frac{\delta {\cal W}^{[0]}}{\delta L_a(x)}
       = \frac{\delta \Gamma^{[0]}}{\delta L_a(x)} \,, \ \ \ \ \ \
\\
&& \frac{\delta {\cal W}^{[0]}}{\delta n_{\alpha}(x) }
       = \frac{\delta \Gamma^{[0]}}{\delta n_{\alpha}(x) } \,, \ \ \ \ \ \
 \frac{\delta {\cal W}^{[0]}}{\delta l_{\alpha}(x) }
       = \frac{\delta \Gamma^{[0]}}{\delta l_{\alpha}(x) } \,, \ \ \ \ \ \
 \frac{\delta {\cal W}^{[0]}}{\delta p_{\alpha}(x)}
       = \frac{\delta \Gamma^{[0]}}{\delta p_{\alpha}(x)} \,, \ \ \ \ \ \
\\
&& \frac{\delta {\cal W}^{[0]}}{\delta n'_{\alpha}(x) }
       = \frac{\delta \Gamma^{[0]}}{\delta n'_{\alpha}(x) } \,, \ \ \ \ \ \
 \frac{\delta {\cal W}^{[0]}}{\delta l'_{\alpha}(x) }
       = \frac{\delta \Gamma^{[0]}}{\delta l'_{\alpha}(x) } \,, \ \ \ \ \ \
 \frac{\delta {\cal W}^{[0]}}{\delta p'_{\alpha}(x)}
       = \frac{\delta \Gamma^{[0]}}{\delta p'_{\alpha}(x)} \,. \ \ \ \ \ \
\end{eqnarray*}
\par
  In order to find the Slavnov--Taylor identity satisfied by the generating
functional for the regular vertex functions, we change the variables in the
path integral of ${\cal Z}^{[0]}$ as follows
\begin{eqnarray*}
&& W_a^{\mu}(x) \rightarrow
   W_a^{\mu}(x) + \delta \zeta \Delta W_a^{\mu}(x)\,, \ \ \ \ \
 W_y^{\mu}(x) \rightarrow
   W_y^{\mu}(x) + \delta \zeta \Delta W_y^{\mu}(x)\,,
\\
&& C_a(x) \rightarrow
    C_a(x) + \delta \zeta \Delta C_a(x) \,, \ \ \ \ \
 C_y(x) \rightarrow C_y(x) \,,
\\
&& \overline{C}_a(x) \rightarrow  \overline{C}_a(x)
                           - \delta \zeta \lambda_a(x)\,,\ \ \ \ \
 \overline{C}_y(x) \rightarrow  \overline{C}_y(x)
                           - \delta \zeta \lambda_y(x)\,, 
\\
&& \psi(x) \rightarrow \psi(x) +\delta \zeta \Delta \psi(x) \,, \ \ \ \ \
 \overline{\psi}(x) \rightarrow \overline{\psi}(x)
        + \delta \zeta \Delta \overline{\psi}(x) \,,
\\
&& \lambda_a(x) \rightarrow  \lambda_a(x) \,, \ \ \ \ \
 \lambda_y(x) \rightarrow  \lambda_y(x) \,.
\end{eqnarray*}
The the changes in $I_s$ and ${\cal L}_{WM}$ lead to
\begin{eqnarray}
&& \int d^4x \Big\{ 
\frac{\delta \Gamma^{[0]}} {\delta K^a_{\mu}(x)}
 \frac{\delta \Gamma^{[0]}} {\delta \widetilde{W}_a^{\mu}(x)} 
+\frac{\delta \Gamma^{[0]}} {\delta K^y_{\mu}(x)}
 \frac{\delta \Gamma^{[0]}} {\delta \widetilde{W}_y^{\mu}(x)}
+ \frac{\delta \Gamma^{[0]}} {\delta L_a(x)}
 \frac{\delta \Gamma^{[0]}} {\delta \widetilde{C}_a(x)} \nonumber\\
&&\ \ \ \ \ \
+\frac{\delta \Gamma^{[0]}} {\delta \widetilde{\nu}_{L \alpha}(x)}
  \frac{\delta \Gamma^{[0]}} {\delta n_{\alpha}(x)} 
+ \frac{\delta \Gamma^{[0]}} {\delta \widetilde{e}_{L \alpha}(x)}
  \frac{\delta \Gamma^{[0]}} {\delta l_{\alpha}(x)} 
+ \frac{\delta \Gamma^{[0]}} {\delta \widetilde{e}_{R \alpha}(x)}
  \frac{\delta \Gamma^{[0]}} {\delta p_{\alpha}(x)} \nonumber\\
&&\ \ \ \ \ \
+\frac{\delta \Gamma^{[0]}} {\delta \widetilde{\overline{\nu}}_{L \alpha}(x)}
  \frac{\delta \Gamma^{[0]}} {\delta n'_{\alpha}(x)} 
+ \frac{\delta \Gamma^{[0]}} {\delta \widetilde{\overline{e}}_{L \alpha}(x)}
  \frac{\delta \Gamma^{[0]}} {\delta l'_{\alpha}(x)} 
+ \frac{\delta \Gamma^{[0]}} {\delta \widetilde{\overline{e}}_{R \alpha}(x)}
  \frac{\delta \Gamma^{[0]}} {\delta p'_{\alpha}(x)} \nonumber\\
&&\ \ \ \ \ \
- \widetilde{\lambda}_a(x)
   \frac{\delta \Gamma^{[0]}} {\delta \widetilde{\overline{C}}_a(x)}
  - \widetilde{\lambda}_y(x) 
  + \frac{\delta \Gamma^{[0]}} {\delta \widetilde{\overline{C}}_y(x)}
  - \langle \Delta {\cal L}_{WM}(x) \rangle^{[0]}
  - \langle \Delta {\cal L}_{\psi m}(x) \rangle^{[0]}
 \Big\} = 0 \,,
\end{eqnarray}
where
\begin{eqnarray*}
&& \langle \Delta {\cal L}_{WM}(x) \rangle^{[0]}
= \frac{1}{N{\cal Z}^{[0]}} \int {\cal D}
  [\overline{\psi},\psi,{\cal W}, \overline{C},C]
  \Delta {\cal L}_{WM}(x)
 {\rm exp} \Big\{ {\rm i} \big( I^{[0]}_{{\rm eff}} + I_s \big) \Big\}\,,
\\
&& \langle \Delta {\cal L}_{WM}(x) \rangle^{[0]}
= \frac{1}{N{\cal Z}^{[0]}} \int {\cal D}
  [\overline{\psi},\psi,{\cal W}, \overline{C},C]
  \Delta {\cal L}_{\psi m}(x)
 {\rm exp} \Big\{ {\rm i} \big( I^{[0]}_{{\rm eff}} + I_s \big) \Big\}\,.
\end{eqnarray*}
With the definitions of $\Delta {\cal L}_{WM}(x)$
and $\Delta {\cal L}_{WM}(x)$
\begin{eqnarray*}
\delta_B {\cal L}_{WM}(x) = \delta \zeta \Delta {\cal L}_{WM}(x) \,,
   \ \ \ \ \ \
 \delta_B {\cal L}_{\psi m}(x) = \delta \zeta \Delta {\cal L}_{\psi m}(x) \,,
\end{eqnarray*}
one can write
\begin{eqnarray*}
 \langle \Delta {\cal L}_{WM}(x) \rangle^{[0]}
 &=& M^2 \widetilde{W}_{a \mu}(x)
       \frac{\delta\Gamma^{[0]}}{\delta K^a_{\mu}(x)}
  + M^2 \Big(\frac{g_1}{g}\Big)^2 \widetilde{W}_{y \mu}(x)
       \frac{\delta\Gamma^{[0]}}{\delta K^y_{\mu}(x)} \\
  &&- M^2 \frac{g_1}{g} \widetilde{W}_{y \mu}(x)
       \frac{\delta\Gamma^{[0]}}{\delta K^3_{\mu}(x)}
   - M^2 \frac{g_1}{g} \widetilde{W}_{3 \mu}(x)
       \frac{\delta\Gamma^{[0]}}{\delta K^y_{\mu}(x)}  \,,\\
\langle \Delta {\cal L}_{\psi m}(x) \rangle^{[0]}
 &=& - m \frac{\delta\Gamma^{[0]}}{\delta l'_{\alpha}(x)}
                                       \widetilde{e}_{R \alpha}(x)
   + m \widetilde{\overline{e}}_{L \alpha}(x) 
                     \frac{\delta\Gamma^{[0]}}{\delta p_{\alpha}(x)} \\
&& - m \frac{\delta\Gamma^{[0]}}{\delta p'_{\alpha}(x)}
                                        \widetilde{e}_{L\alpha}(x)
    + m \widetilde{\overline{e}}_{R \alpha}(x)
                      \frac{\delta\Gamma^{[0]}}{\delta l_{\alpha}(x)} \,.
\end{eqnarray*}
Next, from the invariance of the path integral of ${\cal Z}^{[0]}$ with
respect to the translation of the integration variables $\overline{C}_a(x)$,
$\overline{C}_y(x)$, $\lambda_a(x)$ and $\lambda_y(x)$, one can get a set of
auxiliary identities
\begin{eqnarray}
&& \frac{\delta \Gamma^{[0]}}{\delta \widetilde{\overline{C}}_1(x)}
   - \partial_{\mu}\frac{\delta \Gamma^{[0]}} {\delta K^1_{\mu}(x)}
   - g_1 \widetilde{W}_{y \mu}\frac{\delta \Gamma^{[0]}}
         {\delta K^2_{\mu}(x)}
   - g_1 \widetilde{W}_{2 \mu}\frac{\delta \Gamma^{[0]}}
         {\delta K^y_{\mu}(x)} \nonumber\\
&&\ \ \ \ \ \ \ \ \
   + \frac{i}{2} \frac{mg}{M^2} \Big\{
     \widetilde{\overline{\nu}}_{L \alpha}(x)
                      \frac{\delta\Gamma^{[0]}}{\delta p_{\alpha}(x)} 
     + \widetilde{\nu}_{L \alpha}(x)
                      \frac{\delta\Gamma^{[0]}}{\delta p'_{\alpha}(x)} 
     - \widetilde{\overline{e}}_{R \alpha}(x)
                      \frac{\delta\Gamma^{[0]}}{\delta n_{\alpha}(x)} 
     -\widetilde{e}_{R \alpha}(x)
                      \frac{\delta\Gamma^{[0]}}{\delta n'_{\alpha}(x)} 
     \Big\} = 0  \,, \ \ \ \
\\
&& \frac{\delta \Gamma^{[0]}}{\delta \widetilde{\overline{C}}_2(x)}
   - \partial_{\mu}\frac{\delta \Gamma^{[0]}} {\delta K^2_{\mu}(x)}
   + g_1 \widetilde{W}_{y \mu} \frac{\delta \Gamma^{[0]}}
         {\delta K^1_{\mu}(x)}
   + g_1 \widetilde{W}_{1 \mu} \frac{\delta \Gamma^{[0]}}
         {\delta K^y_{\mu}(x)} \nonumber\\
&&\ \ \ \ \ \ \ \ \
   + \frac{1}{2} \frac{mg}{M^2} \Big\{
     \widetilde{\overline{\nu}}_{L \alpha}(x)
                      \frac{\delta\Gamma^{[0]}}{\delta p_{\alpha}(x)} 
     - \widetilde{\nu}_{L \alpha}(x)
                      \frac{\delta\Gamma^{[0]}}{\delta p'_{\alpha}(x)} 
     + \widetilde{\overline{e}}_{R \alpha}(x)
                      \frac{\delta\Gamma^{[0]}}{\delta n_{\alpha}(x)} 
     -\widetilde{e}_{R \alpha}(x)
                      \frac{\delta\Gamma^{[0]}}{\delta n'_{\alpha}(x)} 
     \Big\} = 0  \,, \ \ \ \
\\
&& \frac{\delta \Gamma^{[0]}}{\delta \widetilde{\overline{C}}_3(x)}
   - \partial_{\mu}\frac{\delta \Gamma^{[0]}} {\delta K^3_{\mu}(x)}
   - g_1 \widetilde{W}_{y \mu}\frac{\delta \Gamma^{[0]}}
         {\delta K^3_{\mu}(x)}
   - g_1 \widetilde{W}_{3 \mu}\frac{\delta \Gamma^{[0]}}
         {\delta K^y_{\mu}(x)} \nonumber\\
&&\ \ \ \ \ \ \ \ \
  + \frac{i}{2} \frac{mg}{M^2} \Big\{
     \widetilde{\overline{e}}_{R \alpha}(x)
                      \frac{\delta\Gamma^{[0]}}{\delta l_{\alpha}(x)} 
     + \widetilde{e}_{R \alpha}(x)
                      \frac{\delta\Gamma^{[0]}}{\delta l'_{\alpha}(x)} 
     - \widetilde{\overline{e}}_{L \alpha}(x)
                      \frac{\delta\Gamma^{[0]}}{\delta p_{\alpha}(x)} 
     -\widetilde{e}_{L \alpha}(x)
                      \frac{\delta\Gamma^{[0]}}{\delta p'_{\alpha}(x)} 
     \Big\} = 0  \,, \ \ \ \
\\
&&\frac{\delta \Gamma^{[0]}}{\delta \widetilde{\overline{C}}_y(x)}
   - \partial_{\mu}\frac{\delta \Gamma^{[0]}} {\delta K^y_{\mu}(x)}
   - g \widetilde{W}_{y \mu}\frac{\delta \Gamma^{[0]}}
         {\delta K^3_{\mu}(x)}
   - g \widetilde{W}_{3 \mu}\frac{\delta \Gamma^{[0]}}
         {\delta K^y_{\mu}(x)} = 0  \,,  
\end{eqnarray}      
and
\begin{eqnarray}
 \frac{\delta \Gamma^{[0]}}{\delta \widetilde{\lambda}_a(x)}
                   = \langle \Phi_a(x) \rangle^{[0]}  \,,\ \ \ \ \
  \frac{\delta \Gamma^{[0]}}{\delta \widetilde{\lambda}_y(x)}
                   = \langle \Phi_y(x) \rangle^{[0]}  \,.
\end{eqnarray}      
where
\begin{eqnarray}
&& \langle \Phi_a(x) \rangle^{[0]}
  = \frac{1}{N{\cal Z}^{[0]}} \int {\cal D}
   [\overline{\psi},\psi,{\cal W}, \overline{C},C,\lambda] \Phi_a(x)
   {\rm exp} \Big\{ {\rm i} \big( I^{[0]}_{{\rm eff}} + I_s \big) \Big\}\,,
\\
&& \langle \Phi_y(x) \rangle^{[0]}
  = \frac{1}{N{\cal Z}^{[0]}} \int {\cal D}
   [\overline{\psi},\psi,{\cal W}, \overline{C},C,\lambda] \Phi_y(x)
   {\rm exp} \Big\{ {\rm i} \big( I^{[0]}_{{\rm eff}} + I_s \big) \Big\}\,.
\end{eqnarray}
Let $\widetilde{\Phi}_a(x) $, $\widetilde{\Phi}_y(x)$,
 $\widetilde{{\cal L}}_{WM}$ and $\widetilde{{\cal L}}_{\psi m}$
 be the results obtained from $\Phi_a(x)$, $\Phi_y(x)$, ${\cal L}_{WM}$
 and ${\cal L}_{\psi m}$ by replacing the field functions with the
 classical field functions and define
\begin{eqnarray}
\overline{\Gamma}^{[0]}
 = \Gamma^{[0]} - \int d^4x \Big\{
       \widetilde{\lambda}_a(x)\widetilde{\Phi}_a(x) 
      + \widetilde{\lambda}_y(x)\widetilde{\Phi}_y(x) 
      + \widetilde{{\cal L}}_{WM} + \widetilde{{\cal L}}_{\psi m}
   \Big\} \,,
\end{eqnarray}
Thus, from (4.6)--(4.11), one gets
\begin{eqnarray}
&& \int d^4x \Big\{
\frac{\delta \overline{\Gamma}^{[0]}} {\delta K^a_{\mu}(x)}
 \frac{\delta \overline{\Gamma}^{[0]}} {\delta \widetilde{W}_a^{\mu}(x)} 
+\frac{\delta \overline{\Gamma}^{[0]}} {\delta K^y_{\mu}(x)}
 \frac{\delta \overline{\Gamma}^{[0]}} {\delta \widetilde{W}_y^{\mu}(x)}
+ \frac{\delta \overline{\Gamma}^{[0]}} {\delta L_a(x)}
 \frac{\delta \overline{\Gamma}^{[0]}} {\delta \widetilde{C}_a(x)}
\nonumber \\
&&\ \ \ \ \ \
+\frac{\delta \overline{\Gamma}^{[0]}} {\delta \widetilde{\nu}_{L \alpha}(x)}
  \frac{\delta \overline{\Gamma}^{[0]}} {\delta n_{\alpha}(x)} 
+ \frac{\delta \overline{\Gamma}^{[0]}} {\delta \widetilde{e}_{L \alpha}(x)}
  \frac{\delta \overline{\Gamma}^{[0]}} {\delta l_{\alpha}(x)} 
+ \frac{\delta \overline{\Gamma}^{[0]}} {\delta \widetilde{e}_{R \alpha}(x)}
  \frac{\delta \overline{\Gamma}^{[0]}} {\delta p_{\alpha}(x)}
\nonumber\\
&&\ \ \ \ \ \
+\frac{\delta \overline{\Gamma}^{[0]}}
       {\delta \widetilde{\overline{\nu}}_{L \alpha}(x)}
  \frac{\delta \overline{\Gamma}^{[0]}} {\delta n'_{\alpha}(x)} 
+ \frac{\delta \overline{\Gamma}^{[0]}}
       {\delta \widetilde{\overline{e}}_{L \alpha}(x)}
  \frac{\delta \overline{\Gamma}^{[0]}} {\delta l'_{\alpha}(x)} 
+ \frac{\delta \overline{\Gamma}^{[0]}}
       {\delta \widetilde{\overline{e}}_{R \alpha}(x)}
  \frac{\delta \overline{\Gamma}^{[0]}} {\delta p'_{\alpha}(x)} 
 \Big\} = 0 \,.
\end{eqnarray}
and
\begin{eqnarray}
&&\frac{\delta \overline{\Gamma}^{[0]}}{\delta \widetilde{\lambda}_a(x)}
  = \langle \Phi_a(x) \rangle^{[0]} - \widetilde{\Phi}_a(x)  \,,\ \ \ \ \
  \frac{\delta \overline{\Gamma}^{[0]}}{\delta \widetilde{\lambda}_y(x)}
    = \langle \Phi_y(x) \rangle^{[0]} - \widetilde{\Phi}_y(x) \,, \\
&& \frac{\delta \overline{\Gamma}^{[0]}}
     {\delta \widetilde{\overline{C}}_1(x)}
 - \partial_{\mu}\frac{\delta \overline{\Gamma}^{[0]}} {\delta K^1_{\mu}(x)}
 - g_1 \widetilde{W}_{y \mu}\frac{\delta \overline{\Gamma}^{[0]}}
         {\delta K^2_{\mu}(x)}
 - g_1 \widetilde{W}_{2 \mu}\frac{\delta \overline{\Gamma}^{[0]}}
         {\delta K^y_{\mu}(x)} \nonumber\\
&&\ \ \ \ \ \  
+ \frac{i}{2} \frac{mg}{M^2} \Big\{
     \widetilde{\overline{\nu}}_{L \alpha}(x)
     \frac{\delta\overline{\Gamma}^{[0]}}{\delta p_{\alpha}(x)} 
     + \widetilde{\nu}_{L \alpha}(x)
     \frac{\delta\overline{\Gamma}^{[0]}}{\delta p'_{\alpha}(x)} 
     - \widetilde{\overline{e}}_{R \alpha}(x)
     \frac{\delta\overline{\Gamma}^{[0]}}{\delta n_{\alpha}(x)} 
     -\widetilde{e}_{R \alpha}(x)
     \frac{\delta\overline{\Gamma}^{[0]}}{\delta n'_{\alpha}(x)} 
     \Big\} = 0  \,, \ \ \ \ \ \
\\
&&\frac{\delta \overline{\Gamma}^{[0]}}{\delta \widetilde{\overline{C}}_2(x)}
 - \partial_{\mu}\frac{\delta \overline{\Gamma}^{[0]}} {\delta K^2_{\mu}(x)}
 + g_1 \widetilde{W}_{y \mu} \frac{\delta \overline{\Gamma}^{[0]}}
         {\delta K^1_{\mu}(x)}
 + g_1 \widetilde{W}_{1 \mu} \frac{\delta \overline{\Gamma}^{[0]}}
         {\delta K^y_{\mu}(x)} \nonumber\\
&&\ \ \ \ \ \ 
+ \frac{1}{2} \frac{mg}{M^2} \Big\{
     \widetilde{\overline{\nu}}_{L \alpha}(x)
          \frac{\delta\overline{\Gamma}^{[0]}}{\delta p_{\alpha}(x)} 
     - \widetilde{\nu}_{L \alpha}(x)
          \frac{\delta\overline{\Gamma}^{[0]}}{\delta p'_{\alpha}(x)} 
     + \widetilde{\overline{e}}_{R \alpha}(x)
          \frac{\delta\overline{\Gamma}^{[0]}}{\delta n_{\alpha}(x)} 
     -\widetilde{e}_{R \alpha}(x)
          \frac{\delta\overline{\Gamma}^{[0]}}{\delta n'_{\alpha}(x)} 
     \Big\} = 0  \,, \ \ \ \ \ \
\\
&&\frac{\delta \overline{\Gamma}^{[0]}}{\delta \widetilde{\overline{C}}_3(x)}
 - \partial_{\mu}\frac{\delta \overline{\Gamma}^{[0]}} {\delta K^3_{\mu}(x)} 
 - g_1 \widetilde{W}_{y \mu}\frac{\delta \overline{\Gamma}^{[0]}}
         {\delta K^3_{\mu}(x)}
 - g_1 \widetilde{W}_{3 \mu}\frac{\delta \overline{\Gamma}^{[0]}}
         {\delta K^y_{\mu}(x)} \nonumber\\
&&\ \ \ \ \ \ 
+ \frac{i}{2} \frac{mg}{M^2} \Big\{
     \widetilde{\overline{e}}_{R \alpha}(x)
     \frac{\delta\overline{\Gamma}^{[0]}}{\delta l_{\alpha}(x)} 
     + \widetilde{e}_{R \alpha}(x)
     \frac{\delta\overline{\Gamma}^{[0]}}{\delta l'_{\alpha}(x)} 
     - \widetilde{\overline{e}}_{L \alpha}(x)
     \frac{\delta\overline{\Gamma}^{[0]}}{\delta p_{\alpha}(x)} 
     -\widetilde{e}_{L \alpha}(x)
     \frac{\delta\overline{\Gamma}^{[0]}}{\delta p'_{\alpha}(x)} 
     \Big\} = 0  \,, \ \ \ \ \ \
\\
&&\frac{\delta \overline{\Gamma}^{[0]}}{\delta \widetilde{\overline{C}}_y(x)}
 - \partial_{\mu}\frac{\delta \overline{\Gamma}^{[0]}} {\delta K^y_{\mu}(x)} 
 - g \widetilde{W}_{y \mu}\frac{\delta \overline{\Gamma}^{[0]}}
         {\delta K^3_{\mu}(x)}
 - g \widetilde{W}_{3 \mu}\frac{\delta \overline{\Gamma}^{[0]}}
         {\delta K^y_{\mu}(x)} = 0  \,.
\end{eqnarray}      
    As mentioned earlier our intention to use the generalized form of
the theory containing $\lambda_a$, $\lambda_y$ and their sources 
is to study the Renormalisability of the theory for which such
sources are absent from the generating functional for the Green functions
and therefore $\langle \Phi_a(x) \rangle^{[0]}$ and
$\langle \Phi_y(x) \rangle^{[0]}$ are equal to zero. We now, according to
$(4.11)$, let vanish
$\frac{\delta \Gamma^{[0]}}{\delta \widetilde{\lambda}_a(x)}$ and
$\frac{\delta \Gamma^{[0]}}{\delta \widetilde{\lambda}_a(x)}$ to make
$\langle \Phi_a(x) \rangle^{[0]}$ and $\langle \Phi_y(x) \rangle^{[0]}$
equal to zero. This means
\begin{eqnarray}
 \widetilde{\Phi}_a(x) = 0 \,,\ \ \ \ \
  \widetilde{\Phi}_y(x) = 0 \,,
\end{eqnarray} 
and
\begin{eqnarray}
\frac{\delta \overline{\Gamma}^{[0]}}{\delta \widetilde{\lambda}_a(x)}
    = 0 \,,\ \ \ \ \
  \frac{\delta \overline{\Gamma}^{[0]}}{\delta \widetilde{\lambda}_y(x)}
    = 0 \,.
\end{eqnarray} 
\par
      In the following we will denote by $\overline{\Gamma}^{[0]}
 [\psi,\overline{\psi},W,\overline{C}, C, \lambda,K,L,n,l,p,n',l',p']$
the functional that is obtained from $\overline{\Gamma}^{[0]}
 [\widetilde{\psi},\widetilde{\overline{\psi}},\widetilde{W},
  \widetilde{\overline{C}}, \widetilde{C}, \widetilde{\lambda},K,\cdots]$
by replacing the classical field functions with the usual field functions.
Assume that the dimensional regularization method is used and the
Slavnov--Taylor identity and the auxiliary identities are guaranteed.
 Denote the tree part and one loop part of $\overline{\Gamma}^{[0]}$ by
$\overline{\Gamma}^{[0]}_0$ and $\overline{\Gamma}^{[0]}_1$  respectively.
 $\overline{\Gamma}^{[0]}_0$ is thus the modified action
$\overline{I}^{[0]}_{eff}$ obtained from $I^{[0]}_{eff}$ by excluding the
mass term and $(\lambda_a,\lambda_y)$ terms. From $(4.15)$
and $(4.17)-(4.22)$ one has
\begin{eqnarray}
&& \Phi_a(x) = 0 \,,\ \ \ \ \  \Phi_y(x) = 0 \,,\\
&& \frac{\delta \overline{\Gamma}^{[0]}}{\delta \lambda_a(x)}
    = 0 \,,\ \ \ \ \
  \frac{\delta \overline{\Gamma}^{[0]}}{\delta \lambda_y(x)} = 0 \,,\\
&& \Lambda_{op} \overline{\Gamma}^{[0]}_0 = 0 \,, \nonumber
\end{eqnarray}
and
\begin{eqnarray}
&& \overline{\Gamma}^{[0]}_0 * \overline{\Gamma}^{[0]}_1
  + \overline{\Gamma}^{[0]}_1 * \overline{\Gamma}^{[0]}_0
  = \Lambda_{op} \overline{\Gamma}^{[0]}_1 = 0 \,,\\
&& \Sigma_a(x) \overline{\Gamma}^{[0]} = 0 \,, \ \ \ \ \ \
  \Sigma_y(x) \overline{\Gamma}^{[0]} = 0 \,,
\end{eqnarray}
where $\Lambda_{op}$,$\Sigma_a(x)$ and $\Sigma_y(x)$ are defined by
\begin{eqnarray}
\Lambda_{op} &=&  \int d^4x \Big\{ 
   \frac{\delta \overline{\Gamma}^{[0]}_0} {\delta K^a_{\mu}(x)}
   \frac{\delta } {\delta W_a^{\mu}(x)} 
 + \frac{\delta \overline{\Gamma}^{[0]}_0} {\delta W_a^{\mu}(x)}
   \frac{\delta } {\delta K^a_{\mu}(x)} 
 + \frac{\delta \overline{\Gamma}^{[0]}_0} {\delta K^y_{\mu}(x)}
   \frac{\delta } {\delta W_y^{\mu}(x)}
 + \frac{\delta \overline{\Gamma}^{[0]}_0} {\delta W_y^{\mu}(x)}
   \frac{\delta } {\delta K^y_{\mu}(x)} \nonumber\\
&&+ \frac{\delta \overline{\Gamma}^{[0]}_0} {\delta L_a(x)}
   \frac{\delta } {\delta C_a(x)}
 + \frac{\delta \overline{\Gamma}^{[0]}_0} {\delta C_a(x)} 
   \frac{\delta } {\delta L_a(x)} 
 + \frac{\delta \overline{\Gamma}^{[0]}_0} {\delta \nu_{L \alpha}(x)}
   \frac{\delta } {\delta n_{\alpha}(x)} 
 + \frac{\delta \overline{\Gamma}^{[0]}_0} {\delta n_{\alpha}(x)}
   \frac{\delta } {\delta \nu_{L \alpha}(x)}\nonumber\\
&&+ \frac{\delta \overline{\Gamma}^{[0]}_0}
   {\delta e_{L \alpha}(x)}\frac{\delta } {\delta l_{\alpha}(x)} 
 + \frac{\delta \overline{\Gamma}^{[0]}_0} {\delta l_{\alpha}(x)}
   \frac{\delta } {\delta e_{L \alpha}(x)} 
 + \frac{\delta \overline{\Gamma}^{[0]}_0} {\delta e_{R \alpha}(x)}
   \frac{\delta } {\delta p_{\alpha}(x)} 
 + \frac{\delta \overline{\Gamma}^{[0]}_0} {\delta p_{\alpha}(x)}
   \frac{\delta } {\delta e_{R \alpha}(x)} \nonumber\\
&&+ \frac{\delta \overline{\Gamma}^{[0]}_0}
       {\delta \overline{\nu}_{L \alpha}(x)}
   \frac{\delta } {\delta n'_{\alpha}(x)} 
 + \frac{\delta \overline{\Gamma}^{[0]}_0} {\delta n'_{\alpha}(x)} 
   \frac{\delta } {\delta \overline{\nu}_{L \alpha}(x)} 
 + \frac{\delta \overline{\Gamma}^{[0]}_0}
       {\delta \overline{e}_{L \alpha}(x)}
   \frac{\delta } {\delta l'_{\alpha}(x)} 
 + \frac{\delta \overline{\Gamma}^{[0]}_0} {\delta l'_{\alpha}(x)} 
   \frac{\delta } {\delta \overline{e}_{L \alpha}(x)} \nonumber\\
&&+ \frac{\delta \overline{\Gamma}^{[0]}_0}
       {\delta \overline{e}_{R \alpha}(x)}
   \frac{\delta } {\delta p'_{\alpha}(x)} 
 + \frac{\delta \overline{\Gamma}^{[0]}_0} {\delta p'_{\alpha}(x)}
   \frac{\delta } {\delta \overline{e}_{R \alpha}(x)} \Big\} \,,\\
\Sigma_1(x)&=&  \frac{\delta}{\delta \overline{C}_1(x)}
       - \partial_{\mu}\frac{\delta}{\delta K^1_{\mu}(x)}
       - g_1 W_{y \mu}\frac{\delta}{\delta K^2_{\mu}(x)}
       - g_1 W_{2 \mu}\frac{\delta}{\delta K^y_{\mu}(x)} \nonumber\\
&&+ \frac{i}{2} \frac{mg}{M^2} \Big\{
     \overline{\nu}_{L \alpha}(x) \frac{\delta}{\delta p_{\alpha}(x)} 
     + \nu_{L \alpha}(x) \frac{\delta}{\delta p'_{\alpha}(x)} 
     - \overline{e}_{R \alpha}(x) \frac{\delta}{\delta n_{\alpha}(x)} 
     - e_{R \alpha}(x)
     \frac{\delta}{\delta n'_{\alpha}(x)} \Big\} = 0  \,, \ \ \ \ \ \
\\
\Sigma_2(x) &=&  \frac{\delta}{\delta \overline{C}_2(x)}
       - \partial_{\mu}\frac{\delta}{\delta K^2_{\mu}(x)}
       + g_1 W_{y \mu}\frac{\delta}{\delta K^1_{\mu}(x)}
       + g_1 W_{1 \mu}\frac{\delta}{\delta K^y_{\mu}(x)} \nonumber\\
&&+ \frac{1}{2} \frac{mg}{M^2} \Big\{
     \overline{\nu}_{L \alpha}(x) \frac{\delta}{\delta p_{\alpha}(x)} 
     - \nu_{L \alpha}(x) \frac{\delta}{\delta p'_{\alpha}(x)} 
     + \overline{e}_{R \alpha}(x) \frac{\delta}{\delta n_{\alpha}(x)} 
     - e_{R \alpha}(x)
     \frac{\delta}{\delta n'_{\alpha}(x)} \Big\} = 0  \,, \ \ \ \ \ \
\\
\Sigma_3(x)&=& \frac{\delta}{\delta \overline{C}_3(x)}
       - \partial_{\mu}\frac{\delta}{\delta K^3_{\mu}(x)}
       - g_1 W_{y \mu}\frac{\delta}{\delta K^3_{\mu}(x)}
       - g_1 W_{3 \mu}\frac{\delta}{\delta K^y_{\mu}(x)}\nonumber\\
&&+ \frac{i}{2} \frac{mg}{M^2} \Big\{
   \overline{e}_{R \alpha}(x)\frac{\delta}{\delta l_{\alpha}(x)} 
  + e_{R \alpha}(x) \frac{\delta}{\delta l'_{\alpha}(x)} 
  - \overline{e}_{L \alpha}(x)\frac{\delta}{\delta p_{\alpha}(x)} 
  - e_{L \alpha}(x)
   \frac{\delta}{\delta p'_{\alpha}(x)} \Big\} = 0  \,, \ \ \ \ \ \
\\
\Sigma_y(x)&=& \frac{\delta}{\delta \overline{C}_y(x)}
       - \partial_{\mu}\frac{\delta}{\delta K^y_{\mu}(x)}
       - g W_{y \mu}\frac{\delta}{\delta K^3_{\mu}(x)}
       - g W_{3 \mu}\frac{\delta}{\delta K^y_{\mu}(x)}   \,.
\end{eqnarray}
The meaning of the notation $A*B$ is 
\begin{eqnarray}
&&A*B = \int d^4x \Big\{
\frac{\delta A} {\delta K^a_{\mu}(x)}
 \frac{\delta B} {\delta W_a^{\mu}(x)} 
+\frac{\delta A} {\delta K^y_{\mu}(x)}
 \frac{\delta B} {\delta W_y^{\mu}(x)}
+ \frac{\delta A} {\delta L_a(x)}
 \frac{\delta B} {\delta C_a(x)} \nonumber\\
&&\ \ \ \ \ \ \ \ \ \ \ \ \ \ \ \ \ \ 
+\frac{\delta A} {\delta \nu_{L \alpha}(x)}
  \frac{\delta B} {\delta n_{\alpha}(x)} 
+ \frac{\delta A} {\delta e_{L \alpha}(x)}
  \frac{\delta B} {\delta l_{\alpha}(x)} 
+ \frac{\delta A} {\delta e_{R \alpha}(x)}
  \frac{\delta B} {\delta p_{\alpha}(x)} \nonumber\\
&&\ \ \ \ \ \ \ \ \ \ \ \ \ \ \ \ \ \
+\frac{\delta A} {\delta \overline{\nu}_{L \alpha}(x)}
  \frac{\delta B} {\delta n'_{\alpha}(x)} 
+ \frac{\delta A} {\delta \overline{e}_{L \alpha}(x)}
  \frac{\delta B} {\delta l'_{\alpha}(x)} 
+ \frac{\delta A} {\delta \overline{e}_{R \alpha}(x)}
  \frac{\delta B} {\delta p'_{\alpha}(x)} 
 \Big\}  \,.
\end{eqnarray}
$(4.24)-(4.26)$ are of course satisfied by the finite part and the pole part
of $\overline{\Gamma}^{[0]}_1$. Thus the equations of the pole part
 $\overline{\Gamma}^{[0]}_1$ are
\begin{eqnarray}         
&& \frac{\delta \overline{\Gamma}^{[0]}_1}{\delta \lambda_a(x)}
    = 0 \,,\ \ \ \ \
  \frac{\delta \overline{\Gamma}^{[0]}_1}{\delta \lambda_y(x)}
    = 0 \,, \\
&&\Lambda_{op} \overline{\Gamma}^{[0]}_1 = 0 \,,\\
&& \Sigma_a(x) \overline{\Gamma}^{[0]}_1 = 0 \,, \ \ \ \ \ \
  \Sigma_y(x) \overline{\Gamma}^{[0]}_1 = 0 \,.
\end{eqnarray}
\par
    It is known [3] that when $m=0$ the theory is renormalisable and
$\overline{\Gamma}^{[0]}_{1,div} $ is a combination of 5 independent terms.
Now one can also find the corresponding solutions of equations
 $(4.33)-(4.35)$. These solutions are as follows 
\begin{eqnarray}
&& T_{(1)} = T_{WL} - T_{GL} - T_{CK} + 2 I_m^{(C)}\,, \\
&& T_{(2)} = T_{WY} - T_{GY} - T_{CKY} \,, \\
&& T_{(3)} = T_{CK} + T_{CKY} + T_{nn'} + T_{ll'} + T_{pp'}  \,, \\
&& T_{(4)} = T_{\nu L} + T_{eL} - T_{nn'} - T_{ll'} - I_m^{(C)} \,,\\ 
&& T_{(5)} = T_{eR} - T_{pp'} - I_m^{(C)} \,,
\end{eqnarray}
where
\begin{eqnarray*}
&& T_{GL} = g \frac{\partial \overline{\Gamma}^{[0]}_0}{\partial g} \,,\ \ \ 
T_{GY} = g_1 \frac{\partial \overline{\Gamma}^{[0]}_0}{\partial g_1} \,,\ \ \
I_m^{(C)} = m \frac{\partial \overline{\Gamma}^{[0]}_0}{\partial m}
          = u \frac{\partial \overline{\Gamma}^{[0]}_0}{\partial u} \,,
\\
&& T_{WL} = \int d^4x \Big\{
   W_a^{\mu}(x) \frac{\delta \overline{\Gamma}^{[0]}_0}
                                 {\delta W_a^{\mu}(x)} 
  + L_a(x) \frac{\delta \overline{\Gamma}^{[0]}_0} {\delta L_a(x)}
 \Big\} \,,
\\
&& T_{WY} = \int d^4x 
   W_y^{\mu}(x) \frac{\delta \overline{\Gamma}^{[0]}_0}
                                 {\delta W_y^{\mu}(x)} \,,
\\
&& T_{CK} = \int d^4x \Big\{
   \overline{C}_a(x) \frac{\delta \overline{\Gamma}^{[0]}_0}
                                 {\delta \overline{C}_a(x)} 
 + C_a(x) \frac{\delta \overline{\Gamma}^{[0]}_0}
                           {\delta C_a(x)} 
 + K^a_{\mu}(x) \frac{\delta \overline{\Gamma}^{[0]}_0}
                     {\delta K^a_{\mu}(x)}
 \Big\} \,,
\\
&& T_{CKY} = \int d^4x \Big\{
   \overline{C}_y(x) \frac{\delta \overline{\Gamma}^{[0]}_0}
                                 {\delta \overline{C}_(x)} 
 + C_y(x) \frac{\delta \overline{\Gamma}^{[0]}_0}
                           {\delta C_y(x)} 
 + K^y_{\mu}(x) \frac{\delta \overline{\Gamma}^{[0]}_0}
                     {\delta K^y_{\mu}(x)}
 \Big\} \,,
\\
&& T_{\nu L}= \int d^4x \Big\{
     \nu_{L \alpha}(x) \frac{\delta \overline{\Gamma}^{[0]}_0}
                                   {\delta \nu_{L \alpha}(x)}
     + \overline{\nu}_{L \alpha}(x)
            \frac{\delta \overline{\Gamma}^{[0]}_0}
                 {\delta \overline{\nu}_{L \alpha}(x)} \Big\} \,,
\\
&& T_{eL} = \int d^4x \Big\{
     e_{L \alpha}(x) \frac{\delta \overline{\Gamma}^{[0]}_0}
                                   {\delta e_{L \alpha}(x)}
    + \overline{e}_{L \alpha}(x)
            \frac{\delta \overline{\Gamma}^{[0]}_0}
                 {\delta \overline{e}_{L \alpha}(x)} \Big\} \,,
\\
&& T_{eR} = \int d^4x \Big\{
      e_{R \alpha}(x) \frac{\delta \overline{\Gamma}^{[0]}_0}
                                   {\delta e_{R \alpha}(x)}
     + \overline{e}_{R \alpha}(x)
            \frac{\delta \overline{\Gamma}^{[0]}_0}
                 {\delta \overline{e}_{R \alpha}(x)} \Big\} \,,
\\
&& T_{nn'} = \int d^4x \Big\{
          n_{\alpha}(x) \frac{\delta \overline{\Gamma}^{[0]}_0}
                                   {\delta n_{\alpha}(x)}
          + n'_{\alpha}(x) \frac{\delta \overline{\Gamma}^{[0]}_0}
                                   {\delta n'_{\alpha}(x)}  \Big\} \,,
\\
&& T_{ll'} = \int d^4x \Big\{
           l_{\alpha}(x) \frac{\delta \overline{\Gamma}^{[0]}_0}
                                     {\delta l_{\alpha}(x)}
           + l'_{\alpha}(x) \frac{\delta \overline{\Gamma}^{[0]}_0}
                                     {\delta l'_{\alpha}(x)} \Big\} \,,
\\
&& T_{pp'} = \int d^4x \Big\{
            p_{\alpha}(x) \frac{\delta \overline{\Gamma}^{[0]}_0}
                                      {\delta p_{\alpha}(x)}
            + p'_{\alpha}(x) \frac{\delta \overline{\Gamma}^{[0]}_0}
                                   {\delta p'_{\alpha}(x)} \Big\}  \,,
\end{eqnarray*}
where the parameter $u$ appearing in the expression of $I_m^{(C)}$ stands
for $m/M^2$. Similar to the case of Ref. [3], $T_{(3)}$ is
$2\big(\overline{\Gamma}^{[0]}_0 -I_{WL}-I_{WY}-I_{\psi}-I_{\psi W}\big)$.
$T_{(1)}$ is a combination of $I_{WL}$, $T_{(3)}$ and
$\int d^4x C_y(x) \frac{\delta \overline{\Gamma}^{[0]}_0}{\delta C_y(x)}$. 
$T_{(2)}$ is a combination of $I_{WY}$ and
$\int d^4x C_y(x) \frac{\delta \overline{\Gamma}^{[0]}_0}{\delta C_y(x)}$.
The sum of $T_{(4)}$ and $T_{(5)}$ is $2\big(I_{\psi}+I_{\psi W}\big)$. 
 $\int d^4x C_y(x) \frac{\delta \overline{\Gamma}^{[0]}_0}{\delta C_y(x)}$
and $T_{(5)}$ can be easily checked to satisfy $(4.33)-(4.35)$. \par
     Since $(4.36)-$$(4.40)$ become the whole independent terms of
$\overline{\Gamma}^{[0]}_{1,div}$ when $m=0$,  a new term that appears when
$m\not= 0$ should include $m$ $\overline{\psi}_{\alpha}$ ${\psi}_{\alpha'}$
as a factor and also satisfies $(4.33)-(4.35)$. First, it is clearly not
possible to form such a solution of $(4.33)-(4.35)$ if negative dimension
coefficients are excluded. Next, if $M^2$ is included as a factor in the
definition of the ghost action so that the ghost fields become dimensionless,
then the modified effective action (without $\lambda$ terms) does not contain
negative dimension coefficients. Thus a new term that can appear in
$\overline{\Gamma}^{[0]}_1$ must be formed with some powers
of $\overline{C_{\sigma}}$ $C_{\sigma'}$ and a factor from
$m \overline{\psi}_{\alpha}$${\psi}_{\alpha'}$, where $\sigma$ and $\sigma'$
stand for $1,2,3,y$. However, $(4.33)-(4.35)$ does not have such a solution
neither. This can easily be seen from $(4.35)$. It follows that
$\overline{\Gamma}^{[0]}_{1,div}$ is a combination of (4.36)--(4.40). Namely
\begin{eqnarray}
\overline{\Gamma}^{[0]}_{1,div}
 =\, \alpha_1^{(1)}T_{(1)} + \alpha_2^{(1)}T_{(2)} + \alpha_3^{(1)}T_{(3)}
    + \alpha_4^{(1)}T_{(4)} + \alpha_5^{(1)} T_{(5)} \,,
\end{eqnarray}
where, $\alpha_1^{(1)}, \cdots, \alpha_5^{(1)}$ are constants of order
 $(\hbar)^1$ and are divergent when the space-time dimension tends to $4$.
\par
    In order to cancel the one loop divergence the counterterm of order
$\hbar^1$ in the action should be chosen as 
\begin{eqnarray}
 \delta I^{[1]}_{count} = - \overline{\Gamma}^{[0]}_{1,div} \,.
\end{eqnarray}
Since
\begin{eqnarray}
 \overline{I}_{eff}^{[0]} = \overline{\Gamma}^{[0]}_0 \,,
\end{eqnarray}
it is known from $(4.41)$ that the sum of $\overline{I}_{eff}^{[0]}$ and
$\delta I^{[1]}_{count}$, to order of $\hbar^1$, can be written as
\begin{eqnarray}
&& \overline{I}_{eff}^{[1]}
  [\psi,\overline{\psi},W,C,\overline{C},
  K,L,n,l,p,n',l',p', g,g_1,u] \nonumber\\
&& \ \ \ \ \ \
  = \overline{I}_{{\rm eff}}^{[0]}[\psi^{[0]},\overline{\psi}^{[0]},
 W^{[0]},C^{[0]},\overline{C}^{[0]}, K^{[0]}, L^{[0]},n^{[0]},n^{'[0]},
   \cdots, g^{[0]}, g_1^{[0]},u^{[0]}]  \,,
\end{eqnarray}
where the bare fields and the bare parameters (to order $(\hbar)^1$) are
defined as
\begin{eqnarray}
&& W^{[0]}_{a \mu} = (Z_3^{[1]})^{1/2} W_{a \mu}
        =\big(1 - \alpha_1^{(1)}\big) W_{a \mu}\,,\ \ \ \
  L^{[0]}_a = (Z_3^{[1]})^{1/2} L_a  \,, 
\\
&&  W^{[0]}_{y \mu} = (Z_3^{'[1]})^{1/2} W_{y \mu}
        = \big(1 - \alpha_2^{(1)}\big) W_{y \mu}\,,
\\
&& C^{[0]}_a = (\widetilde{Z}_3^{[1]})^{1/2}C_a
       = \big( 1 - \alpha_3^{(1)} + \alpha_1^{(1)} \big) C_a \,,
\\
&& \overline{C}^{[0]}_a = (\widetilde{Z}_3^{[1]})^{1/2} \overline{C}_a \,,
 \ \ \ \ \
 K^{a[0]}_{\mu} = (\widetilde{Z}_3^{[1]})^{1/2} K^a_{\mu} \,,
\\
&& C^{[0]}_y = (\widetilde{Z}_3^{'[1]})^{1/2}C_y
     = \big(1 - \alpha_3^{(1)} + \alpha_2^{(1)} \big) C_y \,, 
\\
&&\overline{C}^{[0]}_y = (\widetilde{Z}_3^{'[1]})^{1/2} \overline{C}_y \,,
  \ \ \ \
  K^{y[0]}_{\mu} = (\widetilde{Z}_3^{[1]})^{1/2} K^y_{\mu}\,,
\\
&& \nu_L^{[0]} = (Z_{\nu L}^{[1]})^{1/2} \nu_L
         = \big(1 - \alpha_4^{(1)}\big) \nu_L \,,\ \ \ \
  \overline{\nu}_L^{[0]}
                 = (Z_{\nu L}^{[1]})^{1/2} \overline{\nu}_L \,,
\\
&& e_L^{[0]} = (Z_{eL}^{[1]})^{1/2} e_L
            = (Z_{\nu L}^{[1]})^{1/2}e_L \,,\ \ \ \
   \overline{e}_L^{[0]}
                        = (Z_{eL}^{[1]})^{1/2} \overline{e}_L \,,
\\
&& e_R^{[0]} = (Z_{eR}^{[1]})^{1/2} e_R
        = \big(1 - \alpha_5^{(1)}\big)e_R \,,\ \ \ \
  \overline{e}_R^{[0]}
                     = (Z_{eR}^{[1]})^{1/2} \overline{e}_R  \,,
\\
&&  n^{[0]} = (Z_{(n)}^{[1]})^{1/2} n
      = \big(1 - \alpha_3^{(1)} + \alpha_4^{(1)} \big) n \,,\ \ \ \
    n^{'[0]} = (Z_{(n)}^{[1]})^{1/2} n' \,,
\\
&& l^{[0]}  = (Z_{(l)}^{[1]})^{1/2} l
           = (Z_{(n)}^{[1]})^{1/2} l\,, \ \ \ \
    l^{'[0]} = (Z_{(l)}^{[1]})^{1/2} l' \,,
\\
&&  p^{[0]}  = (Z_{(p)}^{[1]})^{1/2} p
     = \big( 1 - \alpha_3^{(1)} + \alpha_5^{(1)} \big) p \,,\ \ \ \
   p^{'[0]} = (Z_{(p)}^{[1]})^{1/2} p'  \,,
\\
&& g^{[0]} = Z_g^{[1]} g = (Z_3^{[1]})^{-1/2} g \,,\ \ \ \
g_1^{[0]} = Z_g^{'[1]} g_1 = (Z_3^{'[1]})^{-1/2} g_1 \,,  
\\
&& u^{[0]} = \big(1- 2\alpha_1^{(1)}+\alpha_4^{(1)}+\alpha_5^{(1)}\big) u \,.
\end{eqnarray}
\par
     Next let $\Phi_a^{[0]}$ and $\Phi_y^{[0]}$ be obtained from $\Phi_a$
and $\Phi_y$ by replacing the field functions and parameters with the bare
field functions and bare parameters. From $(4.45),(4.46)$ and $(4.57)$ one
has
\begin{eqnarray}
 \Phi_a^{[0]} = (Z_3^{[1]})^{1/2} \Phi_a  \,, \ \ \ \
  \Phi_y^{[0]} = (Z_3^{'[1]})^{1/2} \Phi_y  \,.
\end{eqnarray}
Thus by adding the mass terms and the $\lambda$ terms into
$\overline{I}_{eff}^{[1]}$ and forming
\begin{eqnarray} 
I_{eff}^{[1]} 
 =\, \overline{I}_{eff}^{[1]} + I_{WM} + I_{\psi m}
    + \int d^4x \Big\{
        \lambda_a(x) \Phi_a(x) + \lambda_y(x) \Phi_y(x) \Big\}\,,
\end{eqnarray}
one gets
\begin{eqnarray}
&& I_{eff}^{[1]}
   [\psi,\overline{\psi},W,C,\overline{C}, \lambda,
     K,L,n,l,p,n',l',p', g,g_1, M, m] \nonumber\\
 &&\ \ \ \ \ \ = I_{eff}^{[0]}[\psi^{[0]},\overline{\psi}^{[0]},
      W^{[0]},C^{[0]},\overline{C}^{[0]},\lambda^{[0]},
   K^{[0]}, L^{[0]},n^{[0]},n^{'[0]},
   \cdots, g^{[0]}, g_1^{[0]}, M^{[0]},m^{[0]}] \,,
\end{eqnarray}
where
\begin{eqnarray}
 M^{[0]} = (Z_3^{[1]})^{-1/2} M \,, \ \ \ \
  m^{[0]} = (Z_{eL}^{[1]})^{-1/2} (Z_{eR}^{[1]})^{-1/2} m \,, \ \ \ \
\end{eqnarray}
and
\begin{eqnarray}
 \lambda_a^{[0]} = (Z_3^{[1]})^{-1/2} \lambda_a \,, \ \ \ \
  \lambda_y^{[0]} = (Z_3^{'[1]})^{-1/2} \lambda_y \,.
\end{eqnarray}
\par
Obviously, if the action $I_{eff}^{[1]}$ is used to replace $I_{eff}^{[0]}$
in $(4.3)$ and define  ${\cal Z}^{[1]}$, $\Gamma^{[1]}$ as well as
\begin{eqnarray}
 \overline{\Gamma}^{[1]}
 = \Gamma^{[1]} - I_{WM} -I_{\psi m}- \int d^4x \Big\{ \lambda_a(x)\Phi_a(x) 
                     + \lambda_y(x)\Phi_y(x)  \Big\} \,,
\end{eqnarray}
then one has
\begin{eqnarray}
&&  \overline{\Gamma}^{[1]}[\psi,\overline{\psi},W,C,\overline{C},
  \lambda,K,L,n,l,p,n',l',p', g,g_1, u] \nonumber\\
&&\ \ \ \ \ \ \ \
   = \overline{\Gamma}^{[0]}[\psi^{[0]},\overline{\psi}^{[0]},W^{[0]},
    C^{[0]},\overline{C}^{[0]},\lambda^{[0]},K^{[0]}, L^{[0]},
    n^{[0]},n^{'[0]},\cdots, g^{[0]}, g_1^{[0]}, u^{[0]}] \,.
\end{eqnarray}
From this it is easy to check that, to order $\hbar^1$,
$\overline{\Gamma}^{[1]}$ is finite. Moreover, by changing into bare fields
and bare parameters the fields and parameters in  $(4.15)-$ $(4.22)$  and
then transforming them back into the renormalized fields and renormalized
parameters according to $(4.45)-$$(4.59)$, one can see that, under
condition $(4.23)$, $\overline{\Gamma}^{[1]}$ also satisfies
\begin{eqnarray}
&& \Lambda_{op} \overline{\Gamma}^{[1]} = 0 \,,  \\
&& \frac{\delta \overline{\Gamma}^{[1]}}{\delta \lambda_a(x)}
    = 0 \,,\ \ \ \ \
  \frac{\delta \overline{\Gamma}^{[1]}}{\delta \lambda_y(x)} = 0 \,,
\\
&& \Sigma_a(x) \overline{\Gamma}^{[1]} = 0 \,, \ \ \ \ \ \
  \Sigma_y(x) \overline{\Gamma}^{[1]} = 0 \,.
\end{eqnarray}
\par
     We can now use the inductive method and follow the steps of Ref. [3]
to complete the proof of renormalisability. Assume that up to $n$ loop the
theory has been proved to be renormalisable by introducing the counterterm
$$
 I^{[n]}_{\rm count} = \sum_{l=1}^{n} \delta I^{[l]}_{\rm count}, 
$$
where $\delta I^{[l]}_{count}$ is the counterterm of order $\hbar^l$
and has the form of (4.41),(4.42).
Therefore the modified generating functional $\overline{\Gamma}^{[n]}$ for
the regular vertex, defined by the action 
$$ 
 I^{[n]}_{\rm eff} = I^{[0]}_{\rm eff} + I^{[n]}_{\rm count}, 
$$
satisfied equations $(4.66)-(4.68)$ (under $(4.23)$) and, to order $\hbar^n$,
is finite. This also means that the fields or parameters in each of the
following brackets have the same renormalization factor
$$
 ( W^{[0]}_{a \mu}, L_a), (C_a,\overline{C}_a,K^a_{\mu}),
(C_y,\overline{C}_y,K^y_{\mu}), ({\nu}_L,\overline{\nu}_L,
e_L, \overline{e}_L), (e_R, \overline{e}_R ), (n,n',l,l'), (p,p'),
(\lambda,M,g),
$$
and that
\begin{eqnarray*}
&& Z_g^{'[n]} (Z_3^{'[n]})^{1/2} = 1 \,, \ \ \ \ \
  Z_g^{[n]} (Z_3^{[n]})^{1/2} = 1 \,, \\
&& Z_3^{[n]} \widetilde{Z}_3^{[n]}
= \widetilde{Z}_3^{'[n]} \widetilde{Z}_3^{'[n]}
=  Z_{\nu L}^{[n]} Z_{(n)}^{[n]}
=  Z_{eR}^{[n]} Z_{(p)}^{[n]} \,.
\end{eqnarray*}
\par
    Denote by $\overline{\Gamma}^{[n]}_k$ the part of order $\hbar^{k}$
in $\overline{\Gamma}^{[n]}$. For $k\leq n$, $\overline{\Gamma}^{[n]}_k$ is
equal to $\overline{\Gamma}^{[k]}_k$, because it can not contain the
contribution of a counterterm of order $\hbar^{k+1}$ or higher. Thus
on expanding $\overline{\Gamma}^{[n]}$ to order $\hbar^{n+1}$ one has
$$
\overline{\Gamma}^{[n]}
 = \sum_{k=0}^{n} \overline{\Gamma}^{[k]}_k + \overline{\Gamma}^{[n]}_{n+1}
   + \cdots \,.
$$
Using this and extracting the terms of order $\hbar^{(n+1)}$ from the
equations satisfied by $\overline{\Gamma}^{[n]}$, namely $(4.66)-(4.68)$,
one finds
\begin{eqnarray}
&& \Lambda_{op} \overline{\Gamma}^{[n]}_{n+1} = 0 \,,\\
&& \frac{\delta \overline{\Gamma}^{[n]}_{n+1}}{\delta \lambda_a(x)}
    = 0 \,,\ \ \ \ \
  \frac{\delta \overline{\Gamma}^{[n]}_{n+1}}{\delta \lambda_y(x)} = 0 \,,
\\
&& \Sigma_a(x) \overline{\Gamma}^{[n]}_{n+1} = 0 \,, \ \ \ \ \ \
  \Sigma_y(x) \overline{\Gamma}^{[n]}_{n+1} = 0 \,. 
\end{eqnarray}
Let $\overline{\Gamma}^{[n]}_{n+1,div}$ stand for the pole part of
$\overline{\Gamma}^{[n]}_{n+1}$. By repeating the steps going from $(4.33)$
to $(4.41)$, one can arrive at 
\begin{eqnarray}
\overline{\Gamma}^{[n]}_{n+1,div}
 =\, \alpha_1^{(n+1)} T_{(1)} + \alpha_2^{(n+1)} T_{(2)} 
    + \alpha_3^{(n+1)} T_{(3)} + \alpha_4^{(n+1)} T_{(4)}
    + \alpha_5^{(n+1)} T_{(5)} \,,
\end{eqnarray}
where $\alpha_1^{(n+1)}, \cdots, \alpha_5^{(n+1)}$ are constants of order
$(\hbar)^{n+1}$. Therefore, in order to cancel the  $n+1$ loop divergence
the counterterm of order $\hbar^{n+1}$ should be chosen as
\begin{eqnarray}
 \delta I^{[n+1]}_{count} = - \overline{\Gamma}^{[n]}_{n+1,div}
                      [\psi,\overline{\psi},W,C,\overline{C}] \,.
\end{eqnarray}
Adding this counterterm, the mass term and the $\lambda$ terms to
$\overline{I}_{{\rm eff}}^{[n]}$, one can express the effective action of
 order $\hbar^{n+1}$ as
\begin{eqnarray}
&& I_{{\rm eff}}^{[n+1]}
   [\psi,\overline{\psi},
   W,C,\overline{C},\lambda,K,L,n,l,p,n',l',p', g,g_1, M,m] \nonumber\\
&&\ \ \ \ \ \ \
  = I_{{\rm eff}}^{[0]}[\psi^{[0]},\overline{\psi}^{[0]},W^{[0]},C^{[0]},
    \overline{C}^{[0]},\lambda^{[0]},K^{[0]}, L^{[0]},n^{[0]},n^{'[0]},
   \cdots, g^{[0]}, g_1^{[0]}, M^{[0]}, m^{[0]}] \,,
\end{eqnarray}
where the bare fields and the bare parameters (to order $(\hbar)^{n+1}$) are
defined as
\begin{eqnarray}
&& W^{[0]}_{a \mu} = (Z_3^{[n+1]})^{1/2} W_{a \mu}
        =\big((Z_3^{[n]})^{1/2} - \alpha_1^{(n+1)}\big) W_{a \mu}\,,\ \ \ \
  L^{[0]}_a = (Z_3^{[n+1]})^{1/2} L_a  \,, 
\\
&&  W^{[0]}_{y \mu} = (Z_3^{'[n+1]})^{1/2} W_{y \mu}
        = \big((Z_3^{'[n]})^{1/2} - \alpha_2^{(n+1)}\big) W_{y \mu}\,,
\\
&& C^{[0]}_a = (\widetilde{Z}_3^{[n+1]})^{1/2}C_a
     = \big(( \widetilde{Z}_3^{[n]})^{1/2}
           + (-\alpha_3^{(n+1)} + \alpha_1^{(n+1)}) \big) C_a \,,
\\
&& \overline{C}^{[0]}_a = (\widetilde{Z}_3^{[n+1]})^{1/2} \overline{C}_a \,,
 \ \ \ \ \
 K^{a[0]}_{\mu} = (\widetilde{Z}_3^{[n+1]})^{1/2} K^a_{\mu} \,,
\\
&& C^{[0]}_y = (\widetilde{Z}_3^{'[n+1]})^{1/2}C_y
   = \big((\widetilde{Z}_3^{'[n]})^{1/2}
          + (-\alpha_3^{(n+1)} + \alpha_2^{(n+1)}) \big) C_y \,,
\\
&& \overline{C}^{[0]}_y = (\widetilde{Z}_3^{'[n+1]})^{1/2} \overline{C}_y \,,
  \ \ \ \
  K^{y[0]}_{\mu} = (\widetilde{Z}_3^{[n+1]})^{1/2} K^y_{\mu}\,,
\\
&&  \nu_L^{[0]} = (Z_{\nu L}^{[n+1]})^{1/2} \nu_L
     = \big((Z_{\nu L}^{[n]})^{1/2}
            - \alpha_4^{(n+1)} \big) \nu_L
   \,,\ \ \ \
 \overline{\nu}_L^{[0]}
    = (Z_{\nu L}^{[n+1]})^{1/2} \overline{\nu}_L \,,
\\
&& e_L^{[0]} = (Z_{eL}^{[n+1]})^{1/2} e_L
         = (Z_{\nu L}^{[n+1]})^{1/2} e_L \,,\ \ \ \
   \overline{e}_L^{[0]}
                 = (Z_{eL}^{[n+1]})^{1/2} \overline{e}_L \,,
\\
&& e_R^{[0]} = (Z_{eR}^{[n+1]})^{1/2} e_R
        = \big((Z_{eR}^{[n]})^{1/2}
               - \alpha_5^{(n+1)}\big) e_R \,,\ \ \ \
  \overline{e}_R^{[0]}
                = (Z_{eR}^{[n+1]})^{1/2} \overline{e}_R  \,,
\\
&&  n^{[0]} = (Z_{(n)}^{[n+1]})^{1/2} n
      = \big((Z_{(n)}^{[n]})^{1/2}
             + (-\alpha_3^{(n+1)} + \alpha_4^{(n+1)}) \big) n
      \,,\ \ \ \
    n^{'[0]} = (Z_{(n)}^{[n+1]})^{1/2} n' \,,
\\
&& l^{[0]}  = (Z_{(l)}^{[n+1]})^{1/2} l
           = (Z_{(n)}^{[n+1]})^{1/2} l  \,, \ \ \ \
   l^{'[0]} = (Z_{(l)}^{[n+1]})^{1/2} l' \,,
\\
&&  p^{[0]}  = (Z_{(p)}^{[n+1]})^{1/2} p
     = \big( (Z_{(p)}^{[n]})^{1/2} - \alpha_3^{(n+1)}
                   + \alpha_5^{(n+1)} \big) p \,,\ \ \ \
   p^{'[0]} = (Z_{(p)}^{[n+1]})^{1/2} p'  \,,
\\
&&  g^{[0]} = Z_g^{[n+1]} g = (Z_3^{[n+1]})^{-1/2} g \,,\ \ \ \
   g_1^{[0]} = Z_g^{'[n+1]} g_1 = (Z_3^{'[n+1]})^{-1/2} g_1 \,,
\\
&&  g^{[0]} = Z_g^{[n+1]} g = (Z_3^{[n+1]})^{-1/2} g
     \,,\ \ \ \
   g_1^{[0]} = Z_g^{'[n+1]} g_1  = (Z_3^{'[n+1]})^{-1/2} g_1 \,,
\\
&&  M^{[0]} = Z_M^{[n+1]} M = (Z_3^{[n+1]})^{-1/2} M \,, \ \ \ \
m^{[0]} = Z_m^{[n+1]} m = (Z_{eL}^{[n+1]})^{-1/2}(Z_{eR}^{[n+1]})^{-1/2}M \,,
\end{eqnarray}
and $ \lambda_a^{[0]}, \lambda_y^{[0]} $ are
\begin{eqnarray}
 \lambda_a^{[0]} = (Z_3^{[n+1]})^{-1/2} \lambda_a \,, \ \ \ \
 \lambda_y^{[0]} = (Z_3^{'[n+1]})^{-1/2} \lambda_y \,.
\end{eqnarray}
Therefore, in terms of such bare fields and bare parameters,
 $ \overline{\Gamma}^{[n+1]}$ can be expressed as
\begin{eqnarray}
&& \overline{\Gamma}^{[n+1]}
   [W,C,\overline{C},
   \psi,\overline{\psi},K,L,n,l,p,n',l',p', g,g_1, M] \nonumber\\
&&\ \ \ \ \ \ \
   = \widehat{\Gamma}^{[0]}[W^{[0]},C^{[0]},\overline{C}^{[0]},
   \psi^{[0]},\overline{\psi}^{[0]},K^{[0]}, L^{[0]},n^{[0]},n^{'[0]},
   \cdots, g^{[0]}, g_1^{[0]}, M^{[0]}] \,.
\end{eqnarray}
From this one can conclude that $\overline{\Gamma}^{[n+1]}$, under
$(4.23)$, satisfies $(4.66)-$$(4.68)$ and is finite to order $\hbar^{n+1}$.
That is to say the theory is renormalisable.
\par
\vspace{5mm}
\begin{center}
{\bf V}.\ \ Concluding Remarks 
\end{center}
\par \ \par
    We have expounded that SU$_L$(2) $\times$ U$_Y$(1) electroweak theory
with massive W Z fields and massive electron fields can still be quantized
in a way similar to that used in Ref. [3] by taking into account the
constraint conditions caused by these mass terms and the additional condition chosen
by us. We have also shown that when the $\delta-$ functions appearing in
the path integral of the Green functions and representing the constraint
conditions are rewritten as Fourier integrals with Lagrange multipliers
$\lambda_a$ and $\lambda_y$, the total effective action consisting of the
Lagrange multipliers, ghost fields and the original fields is BRST invariant.
Furthermore, with the help of the  renormalisability  of the theory
without the the mass term of matter fields we have found the general form
of the divergent part of the generating functional $\Gamma$ and proven that
the mass term of the electron fields is also harmless to the
renormalisability of the theory. \par
    It is worth while emphasizing the following special features of
the SU$_L$(2) $\times$ U$_Y$(1) electroweak theory with massive W Z fields
and massive electron fields. (1) These mass terms do not appear in the
divergent part of $\Gamma$. (2) The ghost--electron coupling term
$I_m^{(C)}$, which is caused by the mass term of the electron fields and
contains the negative dimension parameter $m/M^2$, is not an independent
term of the divergent part of $\Gamma$. If this were not the case, the
mass terms would be harmful to the renormalisability of the theory.
\par
    As pointed out in Ref. [3], since the whereabouts of the Higgs Bosons
is still unknown, it is reasonable to ask if the successes of the standard
model of the electroweak theory really depends on the Higgs mechanism and
to pay attention to the theory without the Higgs mechanism.
\par
\vspace{4mm}
\begin{center}
\bf{ACKNOWLEDGMENTS}
\end{center} \par
    We are grateful to Professor Yang Li-ming for helpful discussions. This
work was supported by National Natural Science Foundation of China and
supported in part by Doctoral Programm Foundation of the Institution of
Higher Education of China.
\vspace{4mm}
\par
\ \par
\begin{center}
{\large \bf Refernces}
\end{center}
\par  \noindent
[1]\ Ze-Sen Yang, Zhining Zhou, Yushu Zhong and Xianhui Li,
        hep-th/9912046 7 Dec 1999.
\par  \noindent
[2]\ Ze-Sen Yang, Xianhui Li, Zhining Zhou and Yushu Zhong,
        hep-th/9912034 5 Dec 1999.
\par  \noindent
[3]\ Ze-Sen Yang, Xianhui Li, Zhining Zhou and Yushu Zhong,
       hep-th/0003149 17 Mar 2000.
\par  \noindent
[4]\ M.Carena and C.Wagner, Phys. Rev. {\bf D37}, 560(1988).
\par  \noindent
[5]\ R.Delbourgo and G.Thompson, Phys. Rev. Lett. {\bf 57}, 2610(1986).
\par  \noindent
[6]\ M.Carena and C.Wagner, Phys. Rev. {\bf D37}, 560(1988).
\par  \noindent
[7]\ R.Delbourgo and G.Thompson, Phys. Rev. Lett. {\bf 57}, 2610(1986).
\par  \noindent
[8]\ A.Burnel, Phys. Rev. {\bf D33}, 2981(1986);{\bf D33}, 2985(1986);.
\par  \noindent
[9]\ T.Fukuda, M.Monoa, M.Takeda and K.Yokoyama, \par \ \ \ \ 
Prog. Theor. Phys. {\bf 66},1827(1981);{\bf 67},1206(1982);{\bf 70},284(1983).
\par  \noindent
[10]\ S.L.Glashow, Nucl. Phys. {\bf 22}, 579(1961).
\par  \noindent
[11]\ L.D. Faddeev and V.N. Popov, Phys. Lett. {\bf B25}, 29(1967).
\par  \noindent
[12]\ B.S. De Witt, Phys. Rev. {\bf 162},1195,1239(1967).
\par  \noindent
[13]\ L.D. Faddev and A.A.Slavnov, Gauge Field: Introduction to Quatum Theory,
\par\ \ \ \  The Benjamin Cummings Publishing Company, 1980.
\par  \noindent
[14]\ G.H.Lee and J.H.Yee, Phys. Rev. {\bf D46}, 865(1992). 
\par  \noindent
[15]\ C.Itzykson and F-B.Zuber, Quantum Field Theory, McGraw-Hill,
      New York, 1980.
\par  \noindent
[16]\ Yang Ze-sen, Advanced Quantum Machanics, Peking University Press,
     2-ed.  Beijing, 1995. 
\par
\end{document}